\documentclass[aps,prl,floatfix,twocolumn,showpacs]{revtex4}
\usepackage{epsfig}
\usepackage{amsmath}
\usepackage{amssymb}
\usepackage{amsfonts}
\usepackage{ dsfont }
\usepackage{ comment,color }
\usepackage{lipsum}

\newcommand{\bk}{{{\bf{k}}}}

\newcommand{\bq}{{\bf{q}}}

\newcommand{\beqa}{\begin{eqnarray}}
\newcommand{\eeqa}{\end{eqnarray}}

\newcommand{\ua}{\uparrow}
\newcommand{\da}{\downarrow}

\begin{document}

\hsize\textwidth\columnwidth\hsize\csname@twocolumnfalse\endcsname

\title{
Detecting Topological Superconductivity with 
$\varphi_{0}$ Josephson Junctions}
\author{Constantin Schrade, Silas Hoffman, and Daniel Loss}

\affiliation{Department of Physics, University of Basel,
Klingelbergstrasse 82, CH-4056 Basel, Switzerland}

\date{\today}

\vskip1.5truecm
\begin{abstract}
The interplay of superconductivity, magnetic fields, and spin-orbit interaction 
lies at the heart of topological superconductivity.
Remarkably, the recent experimental discovery of $\varphi_{0}$ Josephson junctions by Szombati {\it et al.} \cite{bib:Kouwenhoven2016}, 
characterized by a finite phase offset in the supercurrent, require the same ingredients as topological superconductors, which suggests  a profound connection between these two distinct phenomena. 
Here, we theoretically show that a quantum dot $\varphi_{0}$ Josephson junction can serve as a new qualitative
indicator for topological superconductivity: Microscopically, we find that the phase
shift in a junction of $s-$wave superconductors is due to the spin-orbit induced mixing of singly occupied states on the qantum dot, while for a topological superconductor junction 
it is due to singlet-triplet mixing. Because of this important difference, 
when the spin-orbit vector of the quantum dot and the external Zeeman field are orthogonal, the $s$-wave superconductors form a $\pi$ Josephson junction while the topological superconductors have a finite offset $\varphi_{0}$ by which topological superconductivity can be distinguished from conventional superconductivity. Our prediction can be immediately tested in nanowire systems currently used for Majorana fermion experiments
and thus offers a new and realistic approach for detecting topological bound states.
\end{abstract}



\maketitle

Non-abelian anyons are the building blocks of topological quantum computers \cite{bib:Nayak2008}.
The simplest realization of a non-abelian anyon are Majorana bound states (MBSs) in topological superconductors (TSs) \cite{bib:Alicea2012}. It has been proposed that such a TS can be induced by 
an $s$-wave superconductor (SC)
in systems of nanowires with spin-orbit interaction (SOI) subject to a Zeeman field \cite{bib:Lutchyn2010,bib:Oreg2010,bib:Klinovaja2012,bib:Kouwenhoven2012}, 
in chains of magnetic atoms
\cite{bib:Yazdani2013,bib:vOppen2013,bib:Perge2014,bib:Meyer2015}
and in topological insulators \cite{bib:Fu2008,bib:Hart2013,bib:Pribiag2015,bib:Wiedenmann2015,bib:Bocquillon2016,bib:Deacon2016}. 
However, providing experimental evidence for the existence of this new phase
of matter has remained a major challenge. 

\begin{figure}[t]  \centering
\includegraphics[width=1\linewidth] {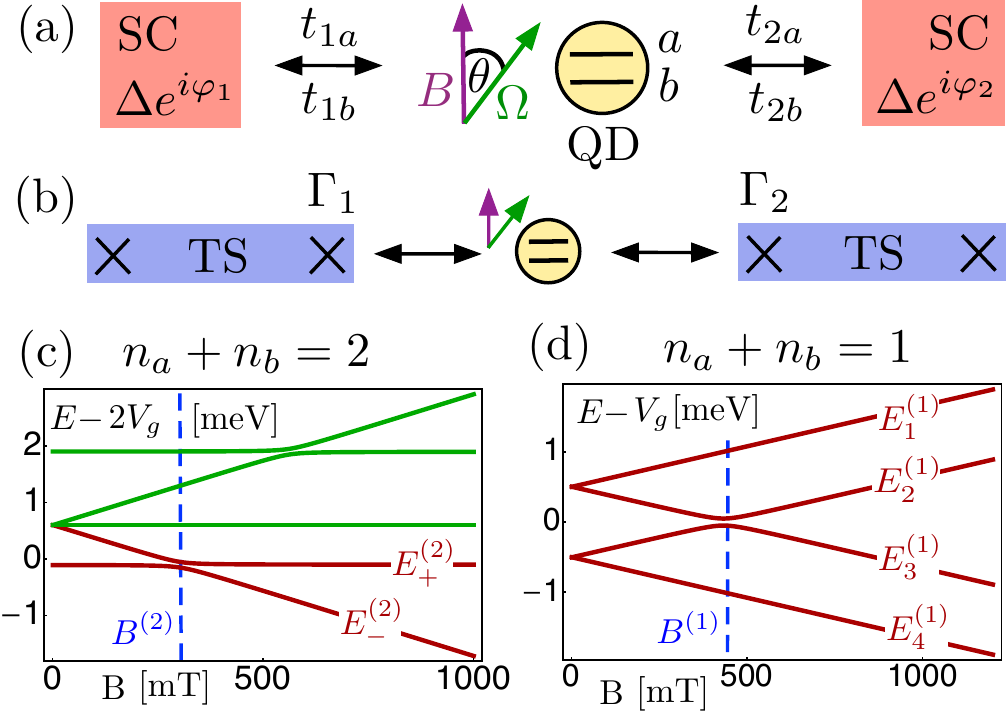}
\caption{
(Color online) 
Setups for $\varphi_{0}$JJs. 
(a) Two $s$-wave SCs (red) are tunnel coupled via a QD (yellow) with two orbitals $a$ and $b$. 
The QD is subject to an external Zeeman field $ B$ at some relative angle $\theta$ to its
SOI axis $\bf\Omega$.  
(b) Same visual encodings. The SCs are replaced by two TSs (blue). The QD now couples to
the two inner MBS (crosses) $\Gamma_{1,2}$ of the TSs. 
(c) Spectrum of the bare QD as a function of $B$ for the double occupancy sector. Red bands contribute to our effective description, green bands do not.  We have chosen $\delta=1$~meV, $g=40$, 
$U=0.9$~meV and $U_{ab}=0.6$~meV, $\Omega=0.1$~meV,
so that $B^{(2)}=302$~mT.
(d) Same as (c) but for the single occupancy sector with $B^{(1)}=  432 $mT.
}\label{fig:1}
\end{figure}

Here we present a new qualitative indicator of MBS based on  $\varphi_{0}$ Josephson junctions ($\varphi_{0}$JJs). In $\varphi_{0}$JJs the Josephson current is offset by a finite phase, $\varphi_0$, so
that a finite supercurrent
flows even when the phase difference between the superconducting leads and the magnetic flux enclosed by the Josephson junction (JJ) vanishes.
Such $\varphi_{0}$JJs have been discussed in systems based on unconventional superconductors 
\cite{bib:Larkin1986,bib:Yip1995,bib:Sigrist1998,bib:Kashiwaya2000,bib:Asano2005}, 
ferromagnets \cite{bib:Buzdin2008,bib:Chan2010,bib:Goldobin2011,bib:Sickinger2012},
quantum point contacts \cite{bib:Reynoso2008}, topological insulators \cite{bib:Dolcini2015}, 
nanowires \cite{bib:Yokoyama2014,bib:Campagnano2015} 
and diffusive systems \cite{bib:Alidoust2013,bib:Bergeret2015}. 
Recently, the connection between $\varphi_{0}$JJs based on nanowires and TSs
has also been discussed \cite{bib:Nesterov2016}. 
Most relevant for the present work, the emergence of a $\varphi_{0}$JJ was theoretically predicted \cite{bib:DellAnna2007,bib:Egger2009,bib:Egger2013} in a system of a quantum dot (QD) with SOI subject to a Zeeman field when coupled to $s$-wave superconducting leads and observed in recent experiments \cite{bib:Kouwenhoven2016}.
Interestingly, the ingredients for observing a
$\varphi_{0}$JJ in this type of system largely overlap with those required to generate MBSs.  
In this work, we focus on two models for $\varphi_{0}$JJs based on QDs which, compared to previous studies \cite{bib:DellAnna2007,bib:Egger2009,bib:Egger2013},
are in the singlet-triplet anticrossing regime. 
In the first
model, two $s$-wave SCs are tunnel coupled via a two-orbital QD with SOI and subject to a Zeeman field, see Fig. \ref{fig:1}(a), wherein we find a finite phase shift caused by the SOI-induced mixing of singly occupied states of the QD.
In the second model, replacing the two $s$-wave SCs by two TSs, see Fig. \ref{fig:1}(b), we again find a finite phase shift which results from the singlet-triplet mixing of the doubly occupied QD states. 
When the spin-orbit vector $\bf{\Omega}$ and the magnetic field are orthogonal,
the system is invariant under a composition of time reversal and mirroring in the plane 
perpendicular to $\bf{\Omega}$, under which the superconducting phase goes to opposite itself; because the energy must be invariant under this symmetry, there can be no terms that are odd 
in the superconducting phase difference in the Hamiltonian and thus no non-trivial phase offset
\cite{bib:Chan2010,bib:Flensberg2016}. 
However, unlike the ground state of the SC leads, the ground states of the TS leads transform nontrivially under the above transformations and we thus anticipate a nonzero phase shift. 
Indeed, we show that the 
phase shift $\varphi_{0}$ is equal to $\pi$ for the $s$-wave superconducting leads, while $\varphi_{0}\neq0,\pi$ for the TSs leads, which can, consequently, be used as a new qualitative indicator of MBSs.

{\it Josephson junction models.} 
Our starting point for both of the JJ models outlined above is the Hamiltonian 
\begin{equation}
\label{Hamiltonian}
H_{\nu}=H_{\text{D}}+H_{\nu,\text{L}}+H_{\nu,\text{t}}\, ,
\end{equation}
where $\nu=\text{S},\text{TS}$ corresponds to the model with $s$-wave SC 
leads and TS leads, respectively.
The first term in this expression $H_{\text{D}}=H_{0}+H_{\text{Z}}+H_{\text{SOI}}$ is the Hamiltonian of an isolated QD. 
Here, $H_{0}=\left(V_{g}+\delta/2\right)n_{a}
+
\left(V_{g}-\delta/2\right)n_{b}
+
U/2
\sum_{\tau}
n_{\tau}
(n_{\tau}-1)\nonumber
+
U_{ab}
n_{a}
n_{b}$
describes a QD with two orbitals $\tau=a,b$ at energy difference $\delta>0$ with respect 
to a gate voltage $V_{g}$. The particle number operator of orbital $\tau$ is  $n_{\tau}=\sum_{s}d^{\dag}_{\tau s}d_{\tau s}$ with $d_{\tau s}$ the electron annihilation operator with spin $s=\ua,\da$ quantized along the $z$-axis in orbital $\tau$.
The intraorbital (interorbital) Coulomb interaction strength is $U$ ($U_{ab}$). 
Furthermore, $H_{\text{Z}}=-g\mu_{B}B\sum_{\tau} (d^\dagger_{\tau\uparrow}d_{\tau\uparrow}-d^\dagger_{\tau\downarrow}d_{\tau\downarrow})/2$ describes a Zeeman field
along the $z$-axis of magnitude $B$ with $g$ the electron $g$-factor and $\mu_{B}$ the
Bohr magneton. 
Lastly, $H_{\text{SOI}}=i\boldsymbol{\Omega}/2 \
\cdot\sum_{s,s'}
(
d^{\dag}_{b s}
\boldsymbol{\sigma}_{ss'}
d_{a s'}-\text{H.c.})$
describes the SOI on the QD, where $\boldsymbol{\Omega}=\Omega(\sin\theta,0,\cos\theta)$, in which $\Omega\neq0$, $\theta\in[0,\pi]$ is the angle of the SOI vector with respect to the Zeeman field, and $\boldsymbol{\sigma}$ is the vector of Pauli matrices.

The second term in Eq.~\eqref{Hamiltonian} describes the isolated superconducting leads.
For the first model, $H_{\text{S,L}}=\sum_{{\eta,\bf k}\sigma}E_{{\bf k}}\gamma_{{\eta,\bf k}\sigma}^\dagger \gamma_{{\eta,\bf k}\sigma}$, where $\gamma_{\eta,{\bf k}\sigma}$ is the quasiparticle annihilation operator in SC $\eta=1,2$ with momentum ${\bf k}$, pseudospin $\sigma=\Uparrow,\Downarrow$, and energy $E_{{\bf k}}=\sqrt{\xi_{{\bf k}}^2+\Delta^2}$ with $\Delta$ the superconducting gap and $\xi_\textbf{k}$ the single-electron dispersion relation in the normal metal state. The non-degenerate ground state of the $s$-wave superconductors, $|0_{\eta}\rangle$, is defined so that $\gamma_{{\eta,\bf k}\sigma}| 0_{\eta}\rangle=0$. 
For the second model, we assume that the localization length of the MBS wavefunctions
is much smaller than the length of TSs. We also neglect contributions of bulk quasiparticles
which is valid for energies much smaller than the energy gap. Consequently
the MBSs are at zero energy and
$H_{\text{TS,L}}=0$. As a result, the ground state of the TS leads is four-fold degenerate which, upon choosing a fixed parity subspace, becomes two-fold degenerate. In the following, we consider the odd parity subspace, however, the results for the even parity ground state subspace are identical.

The last term in Eq.~\eqref{Hamiltonian} describes the tunnel coupling between 
the superconducting leads and the QD. For the first model, it is given by 
\begin{equation}
\label{Tunneling1}
H_{\text{S,t}}=
\sum_{\eta\tau} 
\sum_{{\bf k}s} 
t_{\eta\tau} e^{i\varphi_{\eta}/2} \ c^{\dag}_{\eta,{\bf k}s}d_{\tau s} 
+ \text{H.c.}\, ,
\end{equation}
with $c_{\eta,{\bf k}s}$ being the annihilation operator of an electron with momentum ${\bf k}$
and spin $s$ in SC $\eta$. It is related to the quasiparticle operators by 
$ c_{\eta,{\bf k}\uparrow}=u_{{\bf k}}\gamma_{\eta,{\bf k}\Uparrow}^{}+v_{{\bf k}}\gamma_{\eta,-{\bf k}\Downarrow}^{\dagger}$ and $c_{\eta,-{\bf k}\downarrow}=u_{{\bf k}}\gamma_{\eta,-{\bf k}\Downarrow}^{}-v_{{\bf k}}\gamma_{\eta,{\bf k}\Uparrow}^{\dagger}$ with coherence factors 
$u_{{\bf k}}=(1/\sqrt{2})\sqrt{1+\xi_{\bk}/E_{\bk}}$ and $v_{{\bf k}}=(1/\sqrt{2})\sqrt{1-\xi_{\bk}/E_{\bk}}$.
The tunneling Hamiltonian also contains the superconducting phase $\varphi_{\eta}$ of SC $\eta$ and real, spin and momentum-independent tunneling amplitudes $t_{\eta\tau}$.
For the second model, the coupling of the TSs and the QD is given by 
\begin{equation}
\label{Tunneling2}
H_{\text{TS,t}}=\sum_{\eta\tau}\sum_{s}t_{\eta\tau} e^{i\varphi_{\eta}/2} \ \Gamma_{\eta}d_{\tau s} + \text{H.c.}\, ,
\end{equation}
with $\Gamma_{\eta}$ being the MBS in TS $\eta$ which is spatially closest to the QD \cite{bib:Lopez2013}. 
We assume that its partner $\Gamma'_{\eta}$ at the opposite end of the TS does not couple to the QD. 
However,  they form non-local fermionic operators $C_{1}=(\Gamma'_{1}+i\Gamma_{1})/2$ 
and $C_{2}=(\Gamma_{2}+i\Gamma'_{2})/2$. 

We now proceed with a discussion of $H_{\text{D}}$ in the regime of 
$\delta>U>U_{AB}\gg|\Omega|$, which is common in typical experiments \cite{bib:Kouwenhoven2016}.
First, we address the case of a doubly occupied dot, $n_{a}+n_{b}=2$. For $\Omega=0$, 
the spectrum consists of three singlet (triplet) bands which are constant (split) as a function of the Zeeman field. As experimentally observed in \cite{bib:Fasth2007}, for finite $\Omega$ and $\theta$, the singlet and triplet bands anticross, see Fig. \ref{fig:1}(c). 
In all following discussions, we operate the QD in the regime close to the anticrossing of the singlet 
$\left|S\right\rangle=d^{\dag}_{b\ua}d^{\dag}_{b\da}\left|0_{\text{D}}\right\rangle$ 
and the triplet $\left|T\right\rangle=d^{\dag}_{a\ua}d^{\dag}_{b\ua}\left|0_{\text{D}}\right\rangle$
which occurs at the Zeeman field $B^{(2)}=(\delta-U+U_{ab})/g\mu_{B}$. 
Here, $\left|0_{\text{D}}\right\rangle$ is the vacuum state on the dot.
The effective Hamiltonian, valid to lowest order in $\Omega$, which acts in the
two-level subspace spanned by $\left|S\right\rangle$ and $\left|T\right\rangle$ is 
$H^{\text{(2)}}_{\text{ST}}= (2V_{g}-\delta+U) \left|S\rangle\langle S\right|
+
 (2V_{g}+U_{ab}-g\mu_{B}B) \left|T\rangle\langle T\right|
+
\left[ i\Omega\sin(\theta)/2 \left|T\rangle\langle S\right| + \text{H.c.}\right]$.
The spectrum of $H^{\text{(2)}}_{\text{ST}}$ is given by 
$E^{(2)}_{\pm}$ with corresponding orthonormal eigenstates 
\begin{equation}
\label{doublyoccupied}
\left|E^{(2)}_{\pm}\right\rangle=iS_{\pm}\left|S\right\rangle
+T_{\pm}\left|T\right\rangle.
\end{equation}
Here, $S_{\pm}, T_{\pm}$ are real functions of the system parameters, see \cite{bib:supplemental}. 

Second, we discuss the case of a singly occupied dot, $n_{a}+n_{b}=1$. 
For $\Omega=0$, the energy levels for opposite spins split as a function of the Zeeman field. 
For finite $\Omega$ and $\theta$, an energy gap opens up at the crossing point $B^{(1)}=\delta/g\mu_{B}$ of the spin-up band in orbital $a$ and the spin-down band in orbital $b$, see Fig. \ref{fig:1}(d). We will denote the four eigenvalues of the singly occupied sector by $E^{(1)}_{\lambda}$ for $\lambda=1,...,4$. The corresponding orthonormal eigenstates are given by
\begin{equation}
\label{singlyoccupied}
\begin{split}
|E^{(1)}_{\lambda}\rangle=
\sum_{s}
\left(
i A_{\lambda s} d^{\dag}_{a s} + B_{\lambda s} d^{\dag}_{b s} 
\right)\left|0_\text{D}\right\rangle.
\end{split}
\end{equation}
Here, $A_{\lambda s}$,  $B_{\lambda s}$ are real functions of the system parameters, see \cite{bib:supplemental}. The relative imaginary unit in both Eq.~(\ref{doublyoccupied}) and Eq.~(\ref{singlyoccupied}) is due to the SOI.  We adjust the filling and the gate voltage of the QD,
so that its ground state is given by $E^{(2)}_{-}$ while its first excited states are given by 
$E^{(2)}_{+}$ and $E^{(1)}_{\lambda}$ for some fixed $\lambda$. The seperation between $E^{(2)}_{-}$ 
to the states $E^{(1)}_{\lambda'}$ with $\lambda'\neq\lambda$ is assumed to be large, $|E^{(1)}_{\lambda'}-E^{(2)}_{-}|\gg E^{(1)}_{\lambda}-E^{(2)}_{-}$. 
Finally, the remaining occupancy sectors of the QD, whose energies are much larger than the QD-lead coupling, are not relevant for our results and are hence omitted.

{\it Detecting topological superconductivity.}
In order to calculate the superconducting current, we tune the chemical potential of the superconductors close to the $E^{(2)}_{-}$ level. We require for the SC JJ that $
\pi\nu_{F}t_{\eta\tau}t_{\eta'\tau'}\ll E^{(1)}_{\lambda}-E^{(2)}_{-},\Omega\sin(\theta),\Delta$ with $\nu_{F}$ the normal-state density of states of the leads at the Fermi energy
and for the TS JJ that $t_{\eta\tau}\ll E^{(1)}_{\lambda}-E^{(2)}_{-},\Omega\sin(\theta)$, so that in both cases the states $E^{(2)}_{+}$ and $E^{(1)}_{\lambda}$ on the QD serve as virtual tunneling states. 
Our approach is valid for angles $\theta\in[\theta_{c},\pi-\theta_{c}]$ where 
$\theta_{c}$ is a critical angle determined by the conditions above \cite{bib:supplemental}. 
Furthermore, we work in a temperature regime of $k_{B}T\ll E^{(1)}_{\lambda}-E^{(2)}_{-},\Omega\sin(\theta)$.
The effective tunneling Hamiltonian $H_{\text{S,t}}$ ($H_{\text{TS,t}}$) valid up to fourth (second) order in the tunneling amplitudes acting on the ground state of the isolated dot and $s$-wave (odd parity) ground state of the uncoupled leads is
\begin{align}
\label{Eq6}
H^{\text{eff}}_{\nu\text{,t}}
=
\left(E^{0}_{\nu}\cos\varphi_{\nu}+E^{a}_{\nu}\sin\varphi_{\nu}\right) T_{\nu}
+
\widetilde{E}_{\nu},
\end{align}
with $\varphi_{\text{S}}=2\varphi_{\text{TS}}=\varphi_{1}-\varphi_{2}$ and
$T_{\text{S}}=1$, $T_{\text{TS}}=C^{\dagger}_{1}C_{2}+\text{H.c.}=i\Gamma_{2}\Gamma_{1}$  The first term in Eq.~(\ref{Eq6}) arises due to Cooper pair tunneling across the SC JJ or non-local fermion tunneling
across the TS JJ which splits the ground states of the TS leads. The second term is an energy offset, due to processes
for which there is no such transport. 
At zero temperature, the Josephson current, defined by $I_{\nu}=2e\partial_{\varphi}{E_{\nu,\text{GS}}}/\hbar$ with $E_{\nu,\text{GS}}$ the ground state energy of 
the coupled system, is given by
\begin{equation}
\label{Eq7}
\begin{split}
I_{\nu}&=-I_{\nu}^{c}
\sin(\varphi_{\nu}-\varphi_{\nu}^{ 0}) , \quad
\varphi_{\nu}^{ 0}=\arctan(E^{a}_{\nu}/E^{0}_{\nu}),
\end{split}
\end{equation}
where the critical current is $I_{\nu}^{c}=2\kappa_{\nu}e\sqrt{(E^{0}_{\nu})^{2}+(E^{a}_{\nu})^{2}}\text{sgn}(E^{0}_{\nu})/\hbar$. Because in the TS case the ground state is a function of $\varphi$, the sign of the Josephson energy also depends on the phase difference: $\kappa_{\text{TS}}=-1/2$ when $-E^{0}_{\text{TS}}\cos\varphi_{\text{TS}}-E^{a}_{\text{TS}}\sin\varphi_{\nu}+\widetilde{E}_{\text{TS}}$ is the ground state energy and $\kappa_{\text{TS}}=1/2$ otherwise. In the SC case the ground state is independent of $\varphi$ and therefore $\kappa_{\text{S}}=1$. 
\begin{figure}[!h] \centering
\includegraphics[width=1\linewidth] {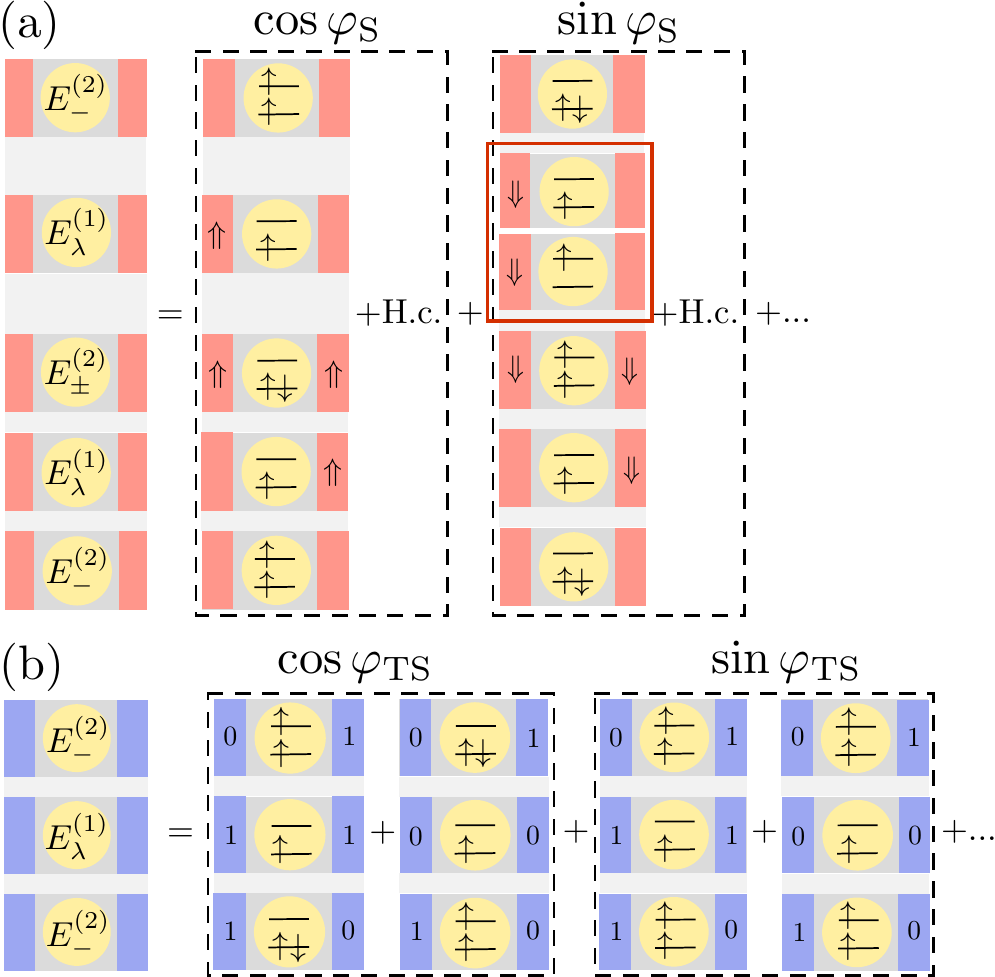}
\caption{(Color online)
(a)
On the left hand side of the equality: the virtual tunneling sequence which leads to $H^{\text{eff}}_{\text{S},t}$ in terms of the eigenstates of the effective dot Hamiltonian. Because the states $|E^{(2)}_{\pm}\rangle$ and $|E^{(1)}_{\lambda}\rangle$ are superpositions of the doubly and singly occupied eigenstates of $H_D$ in the absence of SOI, respectively, $H^{\text{eff}}_{\text{S},t}$ can be written as a sum of the virtual tunneling processes in that basis; two examples of which, contributing to the $\cos\varphi_{\text{S}}$ and $\sin\varphi_{\text{S}}$ terms, are shown on the right side of the equality.
Electron spin (quasiparticle pseudospin) is denoted by $\ua/\da$ ($\Uparrow/\Downarrow$). 
Notice that it is the superposition of \textit{singly} occupied dot states, e.g. in the process 
$|S\rangle\rightarrow d^\dagger_{b\uparrow}|0_D\rangle\rightarrow d^\dagger_{a\uparrow}|0_D\rangle\rightarrow|T\rangle$ (solid red box),
that leads to a finite $\sin\varphi_{\text{S}}$ contribution and therefore a finite $\varphi^0_{\text{S}}$.
(b) Same as (a) but for the case of $H^{\text{eff}}_{\text{TS},t}$. As compared with the SC case, it is the \textit{singlet-triplet} mixing that induces a finite phase shift, e.g. in the contribution to $\cos\varphi_{\textrm{TS}}$. Here, $0$ or $1$ are the eigenvalues of $C_1^\dagger C_1$ and $C_2^\dagger C_2$.
}\label{fig:2}
\end{figure}

\begin{figure} \centering
\includegraphics[width=1\linewidth] {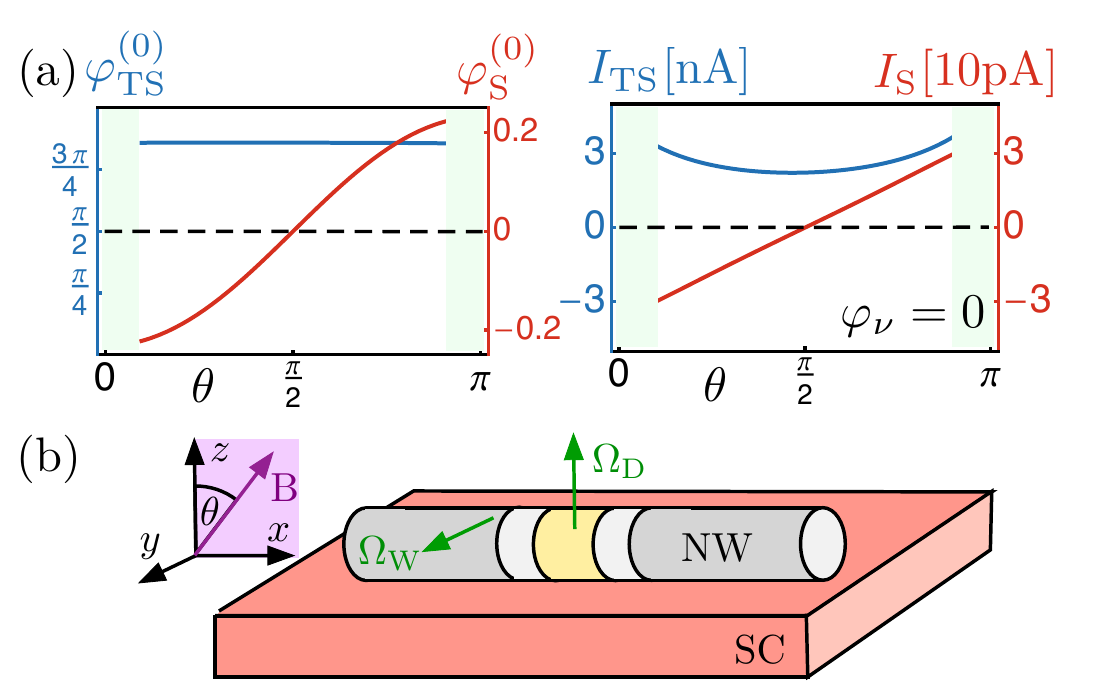}
\caption{(Color online)
(a)
Phase shift $\varphi^{0}_{\nu}(\theta)$ (left panel) and Josephson current $I_{\nu}(\theta)$ at $\varphi_{\nu}=0$ (right panel) for $\lambda=4$ and $\theta\in[\theta_{c},\pi-\theta_{c}]$ with $\theta_{c}=0.3$. System parameters are chosen as in Fig.\ref{fig:1} with
$B=B^{(2)}$, $V_{g}=-0.80$~meV, $t_{1a}=t_{2b}=0.01$~meV, $t_{1b}=0.05$~meV and $t_{2a}=0.04$~meV.
Compared to the SC JJ the phase shift (Josephson current at $\varphi_{\text{S}}=0$) is non-zero for the TS JJ. 
(b) Experimental proposal. A nanowire (dark grey) is proximity coupled to an  $s$-wave SC (red). 
An electric field along the $z$-direction at the SC-wire contact induces a wire SOI axis 
$\bold{\Omega}_{\text{W}}$ along the $y$-direction. An external Zeeman field $\bold{B}$ 
is applied orthogonal to $\bold{\Omega}_{\text{W}}$. 
A QD (yellow) is created by depleting the electron density via gates (light grey regions). 
A backgate contacted to the QD (not shown) induces an electric field along the $y-$direction
axis and hence a SOI axis $\bold{\Omega}_{\text{D}}$ along the $z-$direction. To measure $\varphi^{0}_{\nu}(\theta)$ and $I_{\nu}(\theta)$,
$\bold{B}$ is rotated in the plane orthogonal to $\bold{\Omega}_{\text{W}}$. 
}\label{fig:3}
\end{figure}

Notice that there is a finite phase shift only when $E^{a}_\nu\neq0$. As such, we now turn to a more detailed comparison of the coefficients in Eq.~\eqref{Eq6}. For the BCS JJ, 
\begin{align}
\label{Eq8}
E^{0}_{\text{S}}&=g_{\text{S}} t_{1b}t_{2b}B^{2}_{\lambda \ua}
\left(A^{2}_{\lambda \ua}t_{1b}t_{2b}+B^{2}_{\lambda \ua}t_{1a}t_{2a}\right),\nonumber\\
E^{a}_{\text{S}}&=g_{\text{S}}t_{1b}t_{2b}A_{\lambda \ua}B^{3}_{\lambda \ua}
\left(t_{1a}t_{2b}-t_{1b}t_{2a}\right).
\end{align}
The prefactor $g_{\text{S}}>0$, which is not relevant for the phase shift $\varphi_{\text{S}}^{0}$, includes the coherence factors and energy denominators picked up in the perturbation theory \cite{bib:supplemental}.
Thus, the SC JJ exhibits in general a finite phase shift, when $t_{1a}t_{2b}-t_{1b}t_{2a}\neq0$.
For $\varphi_{S}=0$, the sign of the supercurrent is determined by $\text{sgn}(t_{1a}t_{2b}-t_{1b}t_{2a})$
and $\text{sgn}(A_{\lambda \ua}B_{\lambda \ua})\propto\text{sgn}(\Omega)$.
We now explain the sequence of intermediate states which leads to the contributions in Eq.~(\ref{Eq8}). 
Our initial state on the QD is $E^{(2)}_{-}$. 
To reach the first intermediate state, we remove one electron from the QD, whereupon its state changes to $E^{(1)}_{\lambda}$, and we create an excitation on SC 1 (2). 
Next, we use the superconducting condensate to create an electron on the QD and an excitation 
on SC 2 (1). This changes the QD state to $E^{(2)}_{+}$ \cite{bib:supplemental}.
Third, we return to $E^{(1)}_{\lambda}$ by absorbing one of the dot electrons and the excitation on SC 1 (2) into the condensate. 
Finally, we go back to the initial state 
$E^{(2)}_{-}$ by transferring the excitation on SC 2 (1) back on the QD. 
Because $E^{(1)}_{\lambda}$ is a superposition of different singly occupied QD orbitals,
in the first and third step of this sequence 
the electron on the QD switches orbitals while preserving spin with amplitude $\propto iA_{\lambda\ua}B_{\lambda\ua}$  
while it stays in the same orbital with amplitude $\propto(B_{\lambda\ua})^{2}$ 
or $\propto(A_{\lambda \ua})^{2}$. Thus, the $E^{a}_{\text{S}}$ contribution 
originates from processes in which the electron switches orbitals exactly once, while the remaining processes
yield the $E^{0}_{\text{S}}$ contribution.
The mixing of singlet and triplet states in $E^{(2)}_{\pm}$ gives an overall prefactor, 
which due to the normalization of the states, drops out of Eq.~(\ref{Eq8}). 
Most interestingly, for the case when the relative angle between Zeeman field and SOI axis is 
$\theta=\pi/2$ the phase shift $\varphi_{\text{S}}^{0}$ vanishes, see Fig. \ref{fig:3}(a).
On a microscopic level, this is because now the SOI only mixes opposite spins 
in different orbitals, $A_{\lambda\ua}=B_{\lambda\da}=0$ for $\lambda=1,4$
and
$A_{\lambda\da}=B_{\lambda\ua}=0$ for $\lambda=2,3$
\cite{bib:supplemental}. 
This restricts the number of allowed virtual tunneling processes.
In particular, processes which move the spin between the orbitals without flipping it are prohibited, 
$A_{\lambda \ua}B_{\lambda\ua}=0$ and see Fig. \ref{fig:2}(a).
However, unlike the SC JJ, the TS JJ still allows for nonzero phase shift in that case, see Fig. \ref{fig:3}(a).
At $\theta=\pi/2$, we find that the coefficients in Eq.~\eqref{Eq6} for the TS JJ 
when $\lambda=1,4$ are given by 
\begin{align}
\label{Eq9}
E^{0}_{\text{TS}}(\pi/2)&=g_{\text{TS}}B^{2}_{\lambda\ua}
S_{-}
T_{-}
(t_{1b}t_{2a}-t_{1a}t_{2b}
)\ ,\\
E^{a}_{\text{TS}}(\pi/2)&=-g_{\text{TS}}B^{2}_{\lambda\ua}
\left(
S^{2}_{-}
t_{1b}t_{2b}
+
T^{2}_{-}
t_{1a}t_{2a}
\right)\,,\nonumber
\end{align}
where the prefactor $g_{\text{TS}}>0$ includes the energy denominators of the
perturbation theory \cite{bib:supplemental}. 
In comparison to the SC JJ, the sign of the supercurrent at $\varphi_{\text{TS}}=0$ in the TS JJ 
is determined by parity $i\Gamma_{2}\Gamma_{1}$. If the parity fluctuates, 
the supercurrent exhibits fluctuations as well. So the observation of a phase shift
requires sufficiently long parity life times which can be up to minutes \cite{bib:Kouwenhoven2015}. 
When $\lambda=2,3$ we find that $E^{0}_{\text{TS}}=0$
and $E^{a}_{\text{TS}}\neq0$. For $\lambda=1,4$  we recover the same feature
when $B\gg B^{(2)}$, see Fig. 3 in \cite{bib:supplemental}. In both cases this is the special case of a $\varphi^{0}_{\text{TS}}=\pi/2$ JJ for TS. 
We now focus on the case when $\lambda=1,4$.
Recalling that $E^{(2)}_{-}$ is a superposition of singlet and triplet states, we identify the processes that contribute to Eq.~(\ref{Eq9}): 
$E^{0}_{\text{TS}}(\pi/2)$ comes from virtual tunneling sequences
taking a singlet to a triplet state, with amplitude $\propto iS_{-}T_{-}$, and 
the corresponding sequences taking a triplet to the singlet state, with an amplitude $\propto -iS_{-}T_{-}$. When the order in which the nonlocal fermion is created or destroyed is opposite between these processes, the tunneling sequences differ in phase by $\varphi_{\textrm{TS}}+\pi$ and acquire the same tunneling coefficients so that their sum is proportional to $\cos(\varphi_{\text{TS}})$, see Fig. \ref{fig:2}(b) and \cite{bib:supplemental}.
Distinctly, $E^{a}_{\text{TS}}(\pi/2)$ originates from sequences that take the singlet ($\propto S^{2}_{-}$)
or triplet ($\propto T^{2}_{-}$) to itself. In both cases there exist two sequences that, again, differ in phase by $\varphi_{\textrm{TS}}+\pi$ but have the same tunneling coefficients, so that their sums are $\propto\sin(\varphi_{\text{TS}})$.
{\it Discussion.}
We propose an experiment based on our observation that in general
$\varphi_{\text{S}}^{0}({\pi/2})=0 \quad \text{but} \quad \varphi_{\text{TS}}^{0}({\pi/2})\neq0$.
We consider a nanowire setup similar to \cite{bib:Kouwenhoven2016}, see Fig. \ref{fig:3}(b). 
The wire SOI axis $\bold\Omega_{\text{W}}$, induced by an electric field
along the $z-$axis at the SC-wire contact, is orthogonal to an external Zeeman field 
$\boldsymbol{B}$. Via gating we create a tunnel coupled QD as a short slice in the wire. 
Furthermore we contact the QD to a backgate generating an electric field along the $y-$axis
so that the dot SOI axis $\bold\Omega_{\text{D}}$ is along the $z$-axis. 
We adjust the size of the QD so that the singlet-triplet anticrossing occurs for Zeeman fields
close to the topological phase transition, $g\mu_B B\approx\sqrt{\Delta^{2}+\mu^{2}}$ where $\mu$ is the chemical potential of the SCs and $B=|\boldsymbol{B}|$. Also we adjust the gate voltage $V_{g}$ and the filling of the dot so
that its ground state is $E^{(2)}_{-}$, 
while its first excited states are $E^{(2)}_{+}$ and $E^{(1)}_{4}$. 
Lastly, the chemical potential of the nanowire leads is tuned to $E^{(2)}_{-}$. 
We now position the Zeeman field orthogonal to both $\bold\Omega_{\text{W}}$ and $\bold\Omega_{\text{D}}$. When we now tune the system across the topological phase
transition by varying $B$, we observe a change in the phase shift of the Josephson current from $\pi$ to some non-trivial $\varphi_{0}\neq\pi$. 
Moreover, we can even determine the full dependence of the phase shift 
and Josephson current by rotating $\boldsymbol{B}$ in the plane orthogonal to  $\bold\Omega_{\text{W}}$. 
Interestingly, for typical system parameters of a nanowire QD JJs we find that, at zero phase difference between the leads, $|I_{\text{S}}|\approx10$pA 
while $|I_{\text{TS}}|\approx1$nA, which corresponds to an increase
by three orders of magnitude.

{\it Conclusions.}
We have introduced a new qualitative indicator for the detection of topological superconductivity based on a QD $\varphi_{0}$JJ.
We found that for this setup the trivial SCs always form a $\pi$JJ while the TSs can form a $\varphi_{0}$JJ with $\varphi_{0}\neq 0,\pi$. We have also seen that this change in phase shift is accompanied by
a significant increase in the magnitude of the critical current. 
These observation can be probed by simple modifications of recent 
experimental setups in nanowire QD JJs \cite{bib:Kouwenhoven2016}.

{\it Acknowledgments.}
We acknowledge support  from the Swiss NSF and NCCR QSIT. We are grateful to J. Klinovaja for useful comments.

\begin{widetext}

\newpage

\onecolumngrid

\bigskip 

\begin{center}
\large{\bf Supplemental Material to `Detecting Topological Superconductivity with 
$\varphi_{0}$ Josephson junctions' \\}
\end{center}
\begin{center}
Constantin Schrade$^1$, Silas Hoffman$^1$, and Daniel Loss$^1$
\\
{\it $^1$Department of Physics, University of Basel, Klingelbergstrasse 82, CH-4056 Basel, Switzerland}
\end{center}

\section{A Quantum dot with spin orbit interaction in a Zeeman field}
This first section of the supplemental material provides a more detailed discussion of 
 the model for an isolated QD with 
SOI
subject to an external Zeeman field as given by $H_{\text{D}}$ in the main text. 
The Hilbert space
of the system is spanned by the occupation number states
\begin{equation}
|n_{a\uparrow},n_{a\downarrow}, n_{b\uparrow},n_{b\downarrow}\rangle
=
(d^{\dag}_{a\ua})^{n_{a\uparrow}}
(d^{\dag}_{a\da})^{n_{a\downarrow}}
(d^{\dag}_{b\ua})^{n_{b\uparrow}}
(d^{\dag}_{b\da})^{n_{b\downarrow}}
|0_{\text{D}}\rangle\,,
\end{equation}
where $n_{\tau s}\in\{0,1\}$ is the occupation number of an electron with spin $s$ in orbital $\tau$. 
Since the total number of electrons on the QD is conserved, we can adress each sector
with fixed total occupation number separately. 
\subsection{Double occupancy sector}
We start with an analysis of the double occupancy sector. 
A basis is given by the singlet states 
\begin{equation}
|1,1,0,0\rangle\,, \quad 
|S\rangle=
|0,0,1,1\rangle\,, \quad 
\left(
|1,0,0,1\rangle
-
|0,1,1,0\rangle
\right)/\sqrt{2}\,,
\end{equation}
and the triplet states 
\begin{equation}
|T\rangle=|1,0,1,0\rangle\,, 
\quad 
\left(
|1,0,0,1\rangle
+
|0,1,1,0\rangle
\right)/\sqrt{2}\,,
\quad 
|0,1,0,1\rangle\,.
\end{equation}
Representing $H_{\text{D}}$ in terms of these basis states we find that 
\begin{equation}
\label{Eq4}
H^{(2)}_{\text{D}}=
\begin{pmatrix}
2V_{g}+\delta+U& 0 & 0 & -i \Omega\sin (\theta )/2 &  i \Omega  \cos (\theta )/\sqrt{2} & i \Omega \sin (\theta )/2  \\
0& 2V_{g}-\delta+U  &  0 &-i \Omega  \sin (\theta )/2  & i \Omega  \cos (\theta )/\sqrt{2} & i \Omega    \sin (\theta )/2\\
0 &0 & 2V_{g}+U_{ab} & 0    & 0 &0\\
i \Omega \sin (\theta )/2 & i \Omega  \sin (\theta )/2 & 0 & 2V_{g}+U_{ab}-g\mu_{B}B &0& 0   \\
-i \Omega  \cos (\theta )/\sqrt{2}& -i \Omega  \cos (\theta )/\sqrt{2}&  0 & 0  & 2V_{g}+U_{ab}  &0 \\
-i \Omega   \sin (\theta )/2 &-i \Omega    \sin (\theta )/2  & 0& 0 & 0  & 2V_{g}+U_{ab}+g\mu_{B}B
\end{pmatrix}\,.
\end{equation}
Here, the top left $3\times3$ block acts on the singlet subspace, while the
bottom right $3\times3$ block acts on the triplet subspace and
the off-diagonal blocks contain the SOI which couples the singlet to the triplet subspace.
The spectrum of $H^{(2)}_{\text{D}}$ is depicted in Fig.~1(c) of the main text.
The effective Hamiltonian, valid to lowest order in $\Omega$, which acts in the
two-level subspace spanned by $\left|S\right\rangle$ and $\left|T\right\rangle$ is 
\begin{equation}
H^{\text{(2)}}_{\text{ST}}
=
\begin{pmatrix}
 2V_{g}-\delta+U& -i\Omega \sin(\theta)/2\\
i\Omega \sin(\theta)/2 & 2V_{g}+U_{ab}-g\mu_{B}B
\end{pmatrix}\,.
\end{equation}
It contains the bare energies of the singlet $|S\rangle$ and the triplet $|T\rangle$ on its diagonal.
The SOI interaction then couples these levels via the off-diagonal terms.
The spectrum of $H^{\text{(2)}}_{\text{ST}}$ is given by
\begin{equation}
E^{(2)}_{\pm}
=
2V_{g}+
[(U+U_{ab}-g\mu_{B}B-\delta)/2]
\pm
\sqrt{
\left[
(U-U_{ab}+g\mu_{B}B-\delta)/2
\right]^{2}
+
\left(\Omega\sin(\theta)/2
\right)^{2}
}\,.
\end{equation}
We see that the effect of the SOI is the opening of an
energy gap at the crossing point of the bare singlet and triplet energy levels. 
In terms of the angle between the Zeeman field and the SOI axis, the gap
is maximal when $\theta=\pi/2$ and vanishes when $\theta=0$.
The eigenstates of $H^{\text{(2)}}_{\text{ST}}$ are 
\begin{equation}
\left|E^{(2)}_{\pm}\right\rangle=
\begin{pmatrix}
iS_{\pm} \\
T_{\pm}
\end{pmatrix}
\quad
\Leftrightarrow
\quad 
\left|E^{(2)}_{\pm}\right\rangle=iS_{\pm}\left|S\right\rangle
+T_{\pm}\left|T\right\rangle ,
\end{equation}
where the coefficients are given by
\begin{equation}
\label{Eq7}
\begin{split}
T_{\pm}=\pm\frac{
1
}
{\sqrt{2}}
\sqrt{
1\mp
\frac{
U-U_{ab}+g\mu_{B}B-\delta
}
{
\sqrt{
( U-U_{ab}+g\mu_{B}B-\delta)^{2}
+
\left(\Omega\sin\theta\right)
^{2}
}
}
}\,,
\quad 
S_{-}=-\text{sgn}(\Omega)T_{+}\, 
\quad 
S_{+}=\text{sgn}(\Omega)T_{-}.
\end{split}
\end{equation}
The mixing of the singlet and the triplet is minimal when $\Omega=0$ or $\theta=0$
and it is maximal when $\theta=\pi/2$.

\subsection{Single occupancy sector}
We next discuss the single occupancy sector of the QD which is spanned by the basis states
\begin{equation}
|1,0,0,0\rangle\,, \quad 
|0,1,0,0\rangle\,, \quad 
|0,0,1,0\rangle\,, \quad 
|0,0,0,1\rangle\,.
\end{equation}
The matrix representation of $H_{\text{D}}$ in terms of these basis states is given by
\begin{equation}
\label{HD-Single-Occupancy}
\begin{split}
H^{(1)}_{\text{D}}= 
\frac{1}{2}
\begin{pmatrix}
2V_{g}-\delta-g\mu_{B}B& 0 & i\Omega\cos\theta&i\Omega\sin\theta \\
0& 2V_{g}-\delta+g\mu_{B}B  & i\Omega\sin\theta &-i\Omega\cos\theta  \\
 -i\Omega\cos\theta  & -i\Omega\sin\theta & 2V_{g}+\delta-g\mu_{B}B  &  0   \\
-i\Omega\sin\theta   &   i\Omega\cos\theta& 0& 2V_{g}+\delta+g\mu_{B}B
\end{pmatrix}.
\end{split}
\end{equation}
Here, the top left $2\times 2$ block acts on the subspace of orbital $b$, while the bottom right $2\times 2$ block acts on the subspace of orbital $a$. The off-diagonal blocks contain the SOI 
which couples the $a$ orbital to the $b$ orbital. The spectrum of $H^{(1)}_{\text{D}}$ is depicted in Fig.~1(d)
of the main text and is given by
\begin{equation}
E^{(1)}_{\lambda}=V_{g}+
\frac{1}{2}
\left(
\delta_{\lambda1}
+
\delta_{\lambda2}
-
\delta_{\lambda3}
-
\delta_{\lambda4}
\right)
\sqrt{
\left(\Omega\sin\theta\right)^{2}
+
\left(
g\mu_{B}B
+
\left(
\delta_{\lambda1}
-
\delta_{\lambda2}
-
\delta_{\lambda3}
+
\delta_{\lambda4}
\right)
\sqrt{
\delta^{2}
+
\left(
\Omega\cos\theta
\right)^{2}
}
\right)^{2}
}
.
\end{equation}
Here, $\delta_{\lambda\lambda'}$ for $\lambda,\lambda'=1,...,4$, is the Kronecker delta. 
The eigenstates of $H^{(1)}_{\text{D}}$ are of the form
\begin{equation}
\left|E^{(1)}_{\lambda}\right\rangle=
\begin{pmatrix}
B_{\lambda\ua} \\
B_{\lambda\da} \\
iA_{\lambda\ua} \\
iA_{\lambda\da} 
\end{pmatrix}
\quad
\Leftrightarrow
\quad
|E^{(1)}_{\lambda}\rangle=
\sum_{s}
\left(
i A_{\lambda s} d^{\dag}_{a s} + B_{\lambda s} d^{\dag}_{b s} 
\right)\left|0_\text{D}\right\rangle.
\end{equation}
We now determine the coefficients $A_{\lambda s}$ and $B_{\lambda s}$
for the different relative angles $\theta$ between Zeeman field and SOI axis.

\subsubsection{Zeeman field and SOI axis are orthogonal ($\theta=\pi/2$)}
For $\theta=\pi/2$, the SOI is proportional to $\sigma^{x}$ so that we expect the eigenstates of $H^{(1)}_{\text{D}}$ to be linear combinations of opposite spins in different orbitals.
Indeed, we find that the only coefficients which are non-zero are given by
\begin{equation}
\begin{split}
B_{1\ua}&=A_{4\da}=
\frac{1}{\sqrt{2}}
\sqrt{
1
-
\frac{g\mu_{B}B+\delta}{\sqrt{(g\mu_{B}B+\delta)^{2}+\Omega^{2}}}
}
\,, \quad
B_{4\ua}=-A_{1\da}=
\frac{\text{sgn}(\Omega)}{\sqrt{2}}
\sqrt{
1
+
\frac{g\mu_{B}B+\delta}{\sqrt{(g\mu_{B}B+\delta)^{2}+\Omega^{2}}}
}\,,
\\
A_{3\ua}&=-B_{2\da}=
\frac{1}{\sqrt{2}}
\sqrt{
1
+
\frac{g\mu_{B}B-\delta}{\sqrt{(g\mu_{B}B-\delta)^{2}+\Omega^{2}}}
}
\,, \quad
A_{2\ua}=B_{3\da}=
\frac{\text{sgn}(\Omega)}{\sqrt{2}}
\sqrt{
1
-
\frac{g\mu_{B}B-\delta}{\sqrt{(g\mu_{B}B-\delta)^{2}+\Omega^{2}}}
}\,.
\end{split}
\end{equation}
The remaining coefficients are vanishing, $
B_{1\da}
=
A_{1\ua}
=
A_{2\da}
=
B_{2\ua}
=
A_{3\da}
=
B_{3\ua}
=
B_{4\da}
=
A_{4\ua}=0.
$

\subsubsection{Zeeman field and SOI axis are parallel ($\theta=0,\pi$)}
In the case of $\theta=0,\pi$, the SOI is proportional to $\sigma^{z}$. Consequently, we expect the eigenstates of $H^{(2)}_{\text{D}}$ to be mixtures of same spins in different orbitals.
For $\theta=0$, we find that the non-vanishing coefficients are given by
\begin{equation}
\begin{split}
B_{1\da}=A_{2\da}=-B_{3\ua}=A_{4\ua}
=
\frac{\text{sgn}(\Omega)}{\sqrt{2}}
\sqrt{
1
-
\frac{\delta}{\sqrt{\Omega^{2}+\delta^{2}}}
}
\,, \ \ 
A_{1\da}=-B_{2\da}=A_{3\ua}=B_{4\ua}
=
\frac{1}{\sqrt{2}}
\sqrt{
1
+
\frac{\delta}{\sqrt{\Omega^{2}+\delta^{2}}}
}
\,.
\end{split}
\end{equation}
The remaining coefficients are all zero, $
B_{1\ua}
=
A_{1\ua}
=
A_{2\ua}
=
B_{2\ua}
=
A_{3\da}
=
B_{3\da}
=
B_{4\da}
=
A_{4\da}=0
$. For  $\theta=\pi$, we find find that
\begin{equation}
\begin{split}
B_{1\da}=A_{2\da}=-B_{3\ua}=A_{4\ua}
=
-
\frac{\text{sgn}(\Omega)}{\sqrt{2}}
\sqrt{
1
-
\frac{\delta}{\sqrt{\Omega^{2}+\delta^{2}}}
}
\,, \quad
A_{1\da}=-B_{2\da}=A_{3\ua}=B_{4\ua}
=
\frac{1}{\sqrt{2}}
\sqrt{
1
+
\frac{\delta}{\sqrt{\Omega^{2}+\delta^{2}}}
}
\,.
\end{split}
\end{equation}
As before, the remaining coefficients vanish, $
B_{1\ua}
=
A_{1\ua}
=
A_{2\ua}
=
B_{2\ua}
=
A_{3\da}
=
B_{3\da}
=
B_{4\da}
=
A_{4\da}=0
$. 

\subsubsection{Zeeman field and SOI axis are non-orthogonal and non-parallel ($\theta\neq0,~\pi/2,~\pi$)}
We assume that $\Omega\neq0$; for $\Omega=0$ we note that $H^{(1)}_{\text{D}}$ is already diagonal. 
When $\theta\neq0,~\pi/2,~\pi$, the SOI is proportional to both $\sigma^{x}$ and $\sigma^{z}$. This means that 
the SOI mixes states of all spin species in all orbitals. 
We find that the components of the respective eigenstates are given by 
\begin{equation}
\begin{split}
B_{1\ua}
&=
\frac{1}{N_{1}}
\frac{
g\mu_{B}B+\sqrt{\delta^{2}+\left(\Omega\cos\theta\right)^{2}}
-
\sqrt{
\left(
g\mu_{B}B
+
\sqrt{\delta^{2}+\left(\Omega\cos\theta\right)^{2}}
\right)^{2}
+
\left(\Omega\sin\theta\right)^{2}
}
}
{
\Omega\sin\theta}\,,
\\
B_{2\ua}
&=
\frac{1}{N_{2}}
\frac{
g\mu_{B}B-\sqrt{\delta^{2}-\left(\Omega\cos\theta\right)^{2}}
-
\sqrt{
\left(
g\mu_{B}B-
\sqrt{\delta^{2}+\left(\Omega\cos\theta\right)^{2}}
\right)^{2}
+
\left(\Omega\sin\theta\right)^{2}
}
}
{
\Omega\sin\theta}\,,
\\
B_{3\ua}
&=
\frac{1}{N_{3}}
\frac{
g\mu_{B}B-\sqrt{\delta^{2}-\left(\Omega\cos\theta\right)^{2}}
+
\sqrt{
\left(
g\mu_{B}B-
\sqrt{\delta^{2}+\left(\Omega\cos\theta\right)^{2}}
\right)^{2}
+
\left(\Omega\sin\theta\right)^{2}
}
}{
\Omega\sin\theta}\,,
\\
B_{4\ua}
&=
\frac{1}{N_{4}}
\frac{
g\mu_{B}B+\sqrt{\delta^{2}+\left(\Omega\cos\theta\right)^{2}}
+
\sqrt{
\left(
g\mu_{B}B+\sqrt{\delta^{2}+\left(\Omega\cos\theta\right)^{2}}
\right)^{2}
+
\left(\Omega\sin\theta\right)^{2}
}
}{
\Omega\sin\theta}\,,
\\
B_{1\da}
&=
\frac{1}{N_{1}}
\frac{
\Omega\cos\theta
}
{
\delta+\sqrt{
\delta^{2}+
\left(\Omega\cos\theta\right)^{2}
}
}\quad, \quad
B_{4\da}
=
\frac{1}{N_{4}}
\frac{
\Omega\cos\theta
}
{
\delta+\sqrt{
\delta^{2}+
\left(\Omega\cos\theta\right)^{2}
}
}
\\
\\
B_{2\da}
&=
\frac{1}{N_{2}}
\frac{
\Omega\cos\theta
}
{
\delta-\sqrt{
\delta^{2}+
\left(\Omega\cos\theta\right)^{2}
}
}\quad, \quad
B_{3\da}
=
\frac{1}{N_{3}}
\frac{
\Omega\cos\theta
}
{
\delta-\sqrt{
\delta^{2}+
\left(\Omega\cos\theta\right)^{2}
}
}
\\
A_{\lambda\ua}
&=
\frac{1}{N_{\lambda}}
B_{\lambda\ua}B_{\lambda\da}
\quad, \quad
A_{\lambda\da}
=
\frac{1}{N_{\lambda}},
\end{split}
\end{equation}
where $N_{\lambda}$ is a normalization factors which we choose so that 
$\sqrt{A^{2}_{\lambda\ua}+A^{2}_{\lambda\da}+
B^{2}_{\lambda\ua}+B^{2}_{\lambda\da}}=1$.
The normalization also ensures that when $\theta\rightarrow0,\pi/2,\pi$ the expressions above reproduce the the corresponding limiting cases. 
\\

\section{An s-wave Superconductor $\varphi_{0}$ Josephson junction}
This second section of the supplemental material gives a more detailed discussion 
of the SC JJ described by $H_{\text{S}}$ in the main text. 

\subsection{Effective tunneling Hamiltonian}

\begin{figure}[t]  \centering
\includegraphics[width=\linewidth] {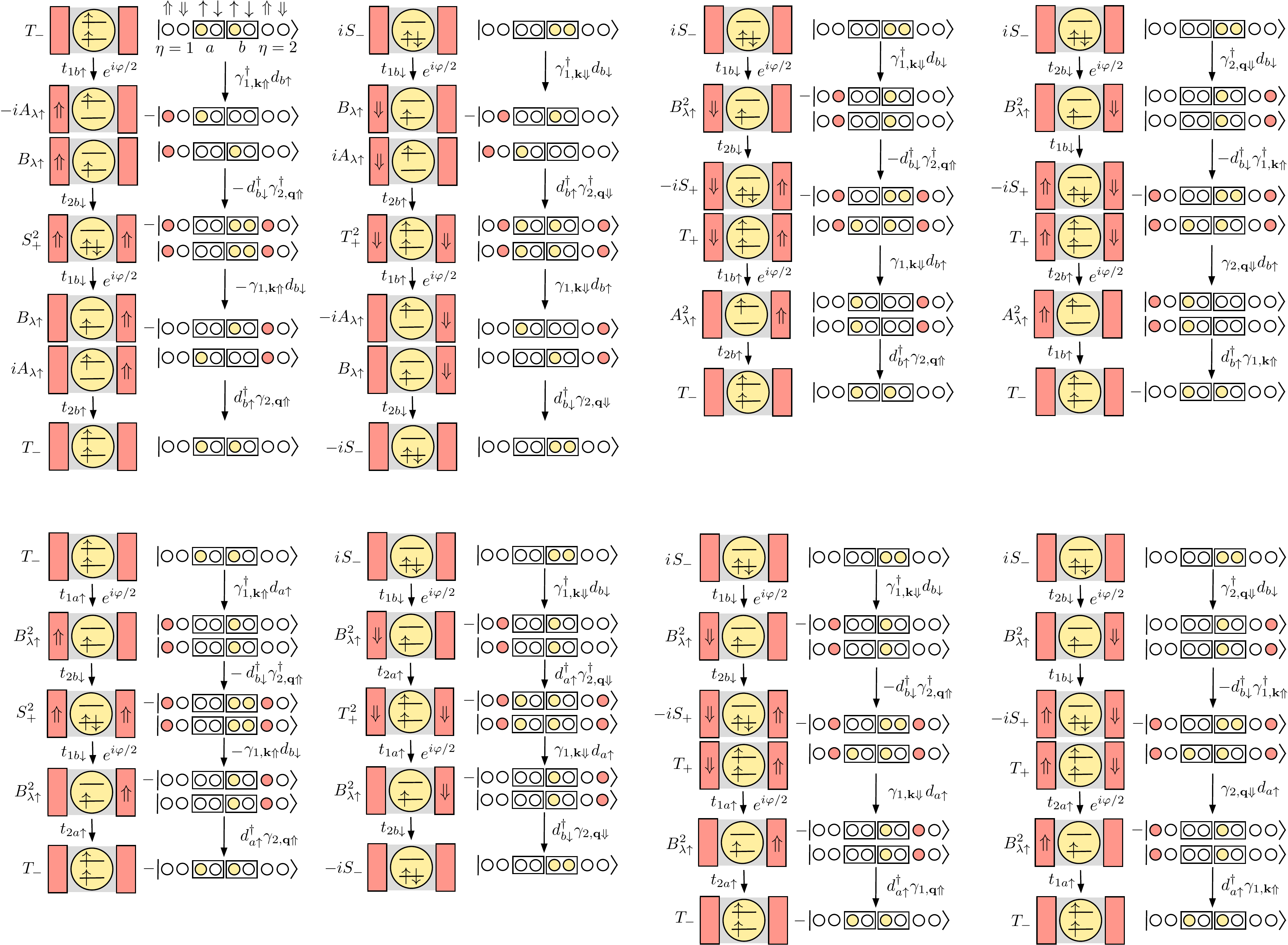}
\caption{
Tunneling sequences (up to hermitian conjugation) of the SC JJ for contributions $\propto\cos\varphi_{\text{S}}$. We use the basis 
$
|n_{1\bk\Uparrow},n_{1\bk\Downarrow},n_{a\uparrow},n_{a\downarrow}, n_{b\uparrow},n_{b\downarrow},n_{2\bq\Uparrow},n_{2\bq\Downarrow}\rangle
=
(\gamma^{\dag}_{1\bk\Uparrow})^{n_{1\bk\Uparrow}}
(\gamma^{\dag}_{1\bk\Downarrow})^{n_{1\bk\Downarrow}}
(d^{\dag}_{a\ua})^{n_{a\uparrow}}
(d^{\dag}_{a\da})^{n_{a\downarrow}}
(d^{\dag}_{b\ua})^{n_{b\uparrow}}
(d^{\dag}_{b\da})^{n_{b\downarrow}}
(\gamma^{\dag}_{2\bq\Uparrow})^{n_{2\bq\Uparrow}}
(\gamma^{\dag}_{2\bq\Downarrow})^{n_{2\bq\Downarrow}}
|0_{1},0_{\text{D}},0_{2}\rangle\,
$. Filled (empty) dots are used to visually represent a filled (an empty) level.
}\label{supp:fig1}
\end{figure}

We begin with a derivation of the effective tunneling Hamiltonian $H^{\text{eff}}_{\text{S,t}}$.
Compared to the main text, 
we allow for a slightly more general tunneling Hamiltonian with spin-dependent
tunneling amplitudes, 
\begin{equation}
\label{Tunneling1}
H_{\text{S,t}}=
\sum_{\eta\tau} 
\sum_{{\bf k}s} 
t_{\eta\tau s} e^{i\varphi_{\eta}/2} \ c^{\dag}_{\eta,{\bf k}s}d_{\tau s} 
+ \text{H.c.}\, 
\end{equation}
Because it is only the relative phase between the two superconductors which is a physical quantity, we assume that $\varphi_{2}=0$ while $\varphi_{1}\equiv\varphi$. 
We now briefly discuss the different tunneling processes which can occur in the system. 
Therefore, we rewrite $H_{\text{S,t}}$
in terms of the quasiparticle operators,
\begin{equation}
\begin{split}
H_{\text{S,t}}
&=
 \sum_{\tau}
\sum_{{\bf k}}
t_{1\tau\ua} e^{i\varphi/2} u_{{\bf k}}\gamma^{\dag}_{1,{\bf k}\Uparrow} d_{\tau\ua} 
+
t_{1\tau\ua} e^{i\varphi/2} 
v_{{\bf k}}\gamma_{1,{\bf k}\Downarrow}
d_{\tau\ua} 
+
t_{2\tau\ua}  u_{{\bf k}}\gamma^{\dag}_{2,{\bf k}\Uparrow} d_{\tau\ua} 
+
t_{2\tau\ua} 
v_{{\bf k}}\gamma_{2,{\bf k}\Downarrow}
d_{\tau\ua} 
\\
&\qquad\quad\ +
t_{1\tau\da} e^{i\varphi/2} 
u_{{\bf k}}\gamma^{\dag}_{1,{\bf k}\Downarrow}
d_{\tau\da} 
-
t_{1\tau\da} e^{i\varphi/2} 
v_{{\bf k}}\gamma_{1,{\bf k}\Uparrow}
d_{\tau\da} 
+
t_{2\tau\da}  
u_{{\bf k}}\gamma^{\dag}_{2,{\bf k}\Downarrow}
d_{\tau\da} 
-
t_{2\tau\da} 
v_{{\bf k}}\gamma_{2,{\bf k}\Uparrow}
d_{\tau\da} 
+\text{H.c.},
\end{split}
\end{equation}
where we have assumed that $\xi_{\bk}=\xi_{-\bk}$.
We see that there are two types of tunneling processes: 
On the one hand, there are processes in which we destroy an electron on the dot and create a
quasiparticle on one of the SC leads (or vice versa). Here, electrons and quasiparticles carry
the same type of spin or pseudospin. 
On the other hand, there are processes in which we use the superconducting condensate
to simultaneously create (or destroy) an electron on the dot and a quasiparticle on the SC leads. 
In this case, electron and quasiparticle always carry the opposite type of spin or pseudospin. 
Because of our convention for the superconducting phases, whenever we destroy (create) 
an electron on the dot and destroy or create a quasiparticle on SC $\eta=1$ we pick up a phase of $e^{i\varphi/2}$ ($e^{-i\varphi/2}$) during the tunneling process.
\begin{figure}[t]  \centering
\includegraphics[width=\linewidth] {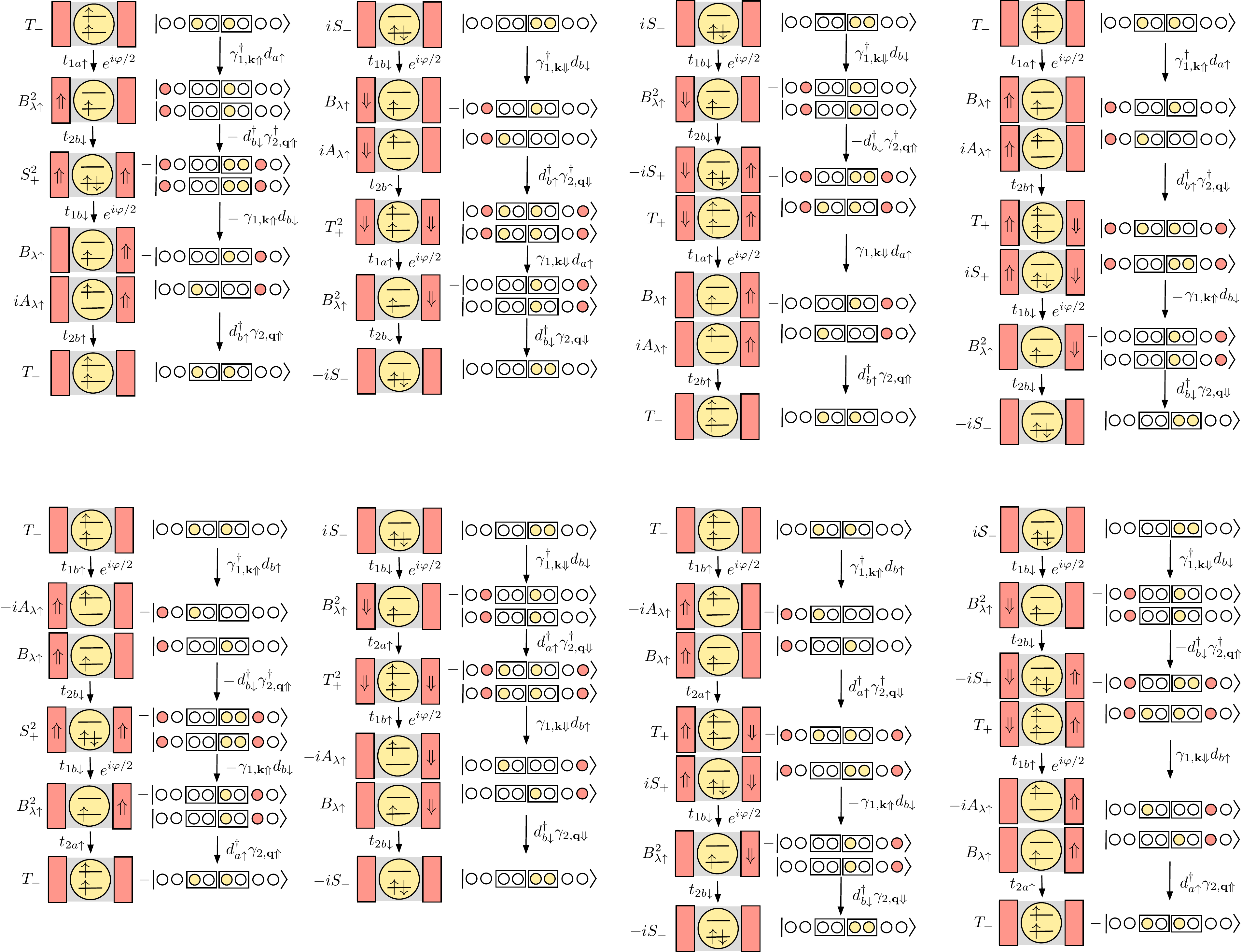}
\caption{
Same as Fig. 1 but for contributions $\propto\sin\varphi_{\text{S}}$ to the effective
Hamiltonian of the SC JJ. }
\label{supp:fig2}
\end{figure}

We now derive the effective tunneling Hamiltonian $H^{\text{eff}}_{\text{S,t}}$ 
using the projection method
\cite{Auerba94}. 
Up to fourth order in the tunneling amplitudes we find that
\begin{equation}
\label{Eq17}
\begin{split}
H^{\text{eff}}_{\text{S,t}}
&= 
P_{\text{S}} H_{\text{S,t}}(E^{(2)}_{-}-H_{\text{D}}-H_{\text{S,L}})^{-1}(1-P_{\text{S}}) H_{\text{S,t}}\, P_{\text{S}} \\ &+
P_{\text{S}}\, H_{\text{S,t}}\left[(E^{(2)}_{-}-H_{\text{D}}-H_{\text{SC,L}})^{-1}(1-P_{\text{S}}) H_{\text{S,t}}\right]^3\, P_{\text{S}}, \; 
\end{split}
\end{equation}
where $P_{\text{S}}=|0_{1},E^{(2)}_{-},0_{2}\rangle\langle0_{1},E^{(2)}_{-},0_{2}|$
is the projector on the $E^{(2)}_{-}$ state on the dot and the ground states of the SC leads. It acts within the reduced Hilbert space of the states $E^{(2)}_{\pm}, E^{(1)}_{\lambda}$ 
on the dot and the full Hilbert space of the SC leads. 
Evaluating Eq.~\eqref{Eq17} yields an expression as given by Eq.~(6) in the main text 
with $\nu=\text{S}$ and
\begin{align}
E^{0}_{\text{S}}&=g_{\text{S}} t_{1b\da}t_{2b\da}B^{2}_{\lambda\ua}
\left(A^{2}_{\lambda\ua}t_{1b\ua}t_{2b\ua}+B^{2}_{\lambda\ua}t_{1a\ua}t_{2a\ua}\right)\nonumber\\
E^{a}_{\text{S}}&=g_{\text{S}}t_{1b\da}t_{2b\da}A_{\lambda\ua}B^{3}_{\lambda\ua}
\left(t_{1a\ua}t_{2b\ua}-t_{1b\ua}t_{2a\ua}\right).
\end{align}
We point out that unlike Eq.~(8) in the main text, this results holds also for spin-dependent tunneling amplitudes. The coupling constant is given by
\begin{equation}
g_{\text{S}}
=
2
\sum_{\bk,\bf{q}}
\frac{
u_{\bk}u_{\bf{q}}
v_{\bk}v_{\bf{q}}
}{
(E^{(1)}_{\lambda}+E_{\bf{q}}-E^{(2)}_{-})
(E^{(2)}_{+}+E_{\bf{k}}+E_{\bf{q}}-E^{(2)}_{-})
(E^{(1)}_{\lambda}+E_{\bf{k}}-E^{(2)}_{-})
}>0.
\end{equation}
We give a complete table of the tunneling sequences (up to hermitian conjugation) contributing to the Cooper pair transport in Fig.~\ref{supp:fig1} and Fig.~\ref{supp:fig2}. 
Here, we note that the sum of the processes in each row of Fig.~\ref{supp:fig1} and Fig.~\ref{supp:fig2}
is $\propto(S_{+}T_{-}-S_{-}T_{+})^{2}$. 
This factor is unity because the states $E^{(2)}_{\pm}$ are orthonormal, see Eq.~\eqref{Eq7}.
This explains why the singlet-triplet mixing does not enter the effective tunneling Hamiltonian.
We omit the presentation of $\widetilde{E}_{\text{S}}$ since it is not relevant
to compute the Josephson current.
The phase shifts $\varphi^{0}_{\text{S}}(\theta)$ and Josephson currents $I_{\text{S}}(\theta)$
 at $\varphi_{\text{S}}=0$ are plotted in Fig.~\ref{supp:fig6}.

\section{A Topological Superconductor $\varphi_{0}$ Josephson junction}

\subsection{Effective tunneling Hamiltonian}

\begin{figure}[t]  \centering
\includegraphics[width=\linewidth] {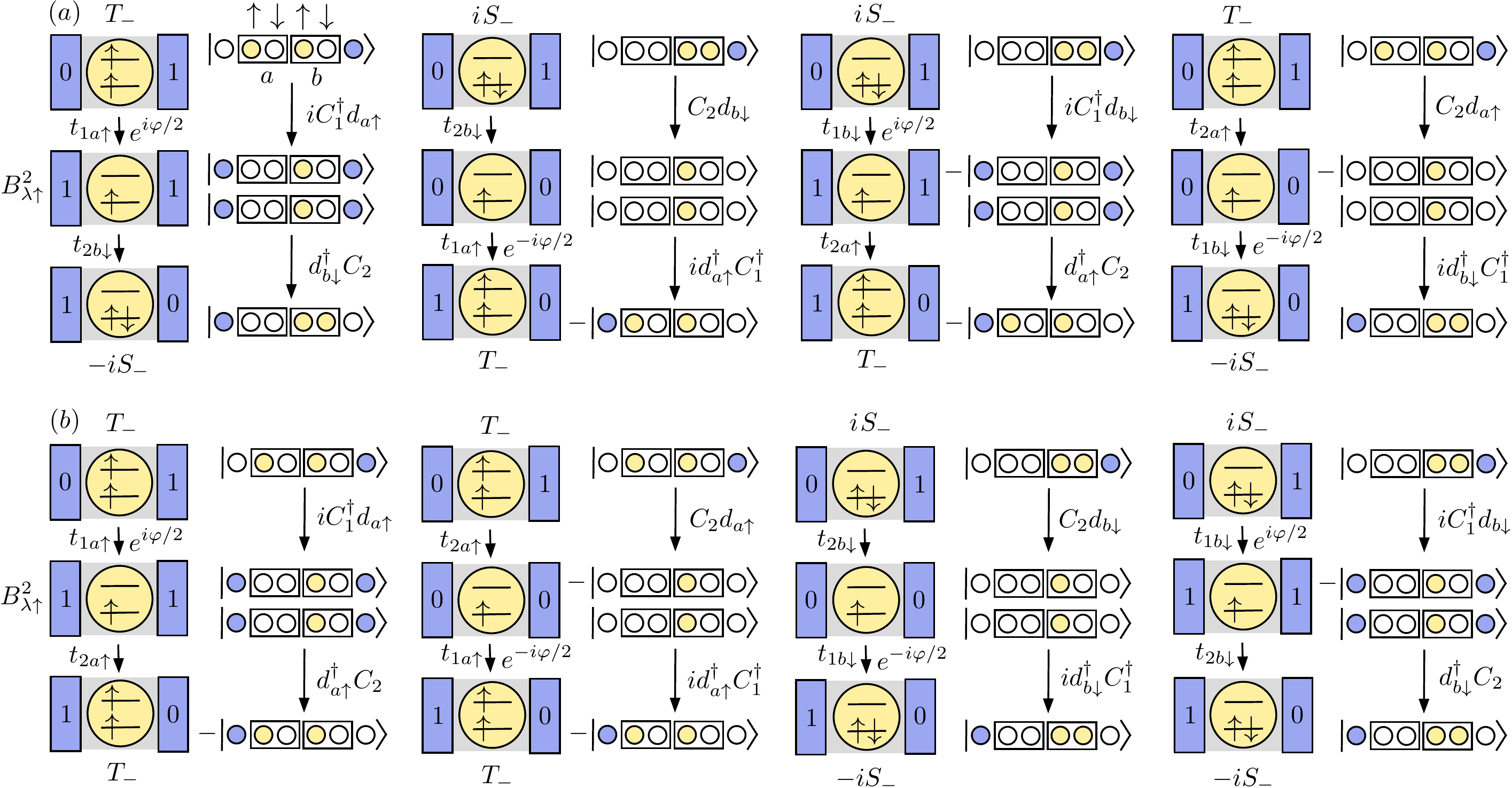}
\caption{
Tunneling sequences of the TS JJ for $\theta=\pi/2$. We use the basis 
$
|n_{1},n_{a\uparrow},n_{a\downarrow}, n_{b\uparrow},n_{b\downarrow},n_{2}\rangle
=
(C^{\dag}_{1})^{n_{1}}
(d^{\dag}_{a\ua})^{n_{a\uparrow}}
(d^{\dag}_{a\da})^{n_{a\downarrow}}
(d^{\dag}_{b\ua})^{n_{b\uparrow}}
(d^{\dag}_{b\da})^{n_{b\downarrow}}
(C^{\dag}_{2})^{n_{2}}
|0_{1},0_{\text{D}},0_{2}\rangle\,
$. Filled (empty) dots are used to visually represent a filled (an empty) level.
(a) Tunneling sequences that give contributions $\propto\cos(\varphi_{\text{TS}})$. 
(b) Tunneling sequences that give contributions $\propto\sin(\varphi_{\text{TS}})$. 
}\label{supp:fig3}
\end{figure}

We devote this third part of the supplemental material to the derivation and discussion of the effective tunneling Hamiltonian $H^{\text{eff}}_{\text{TS,t}}$
for the TS JJ. 
Similar to the SC JJ, we also allow for spin-dependent tunneling amplitudes in the tunneling
Hamiltonian, 
\begin{equation}
\label{Tunneling2}
H_{\text{TS,t}}=\sum_{\eta\tau}\sum_{s}t_{\eta\tau s} e^{i\varphi_{\eta}/2} \ \Gamma_{\eta}d_{\tau s} + \text{H.c.}\, 
\end{equation}
For our derivation we adopt the same assumptions as in the main text.
Compared to the SC JJ 
the lowest order processes which contribute to the Josephson current are of second
order in the tunneling amplitudes. In particular these processes do not mix 
the total fermion parity of the TS leads. Because of that, we focus on the odd parity 
subspace of the TSs. The results for the even parity subspace of the TSs are identical.
The effective tunneling Hamiltonian up to second order in the tunneling amplitudes is given by, 
\begin{equation}
\label{Eq21}
\begin{split}
H^{\text{eff}}_{\text{TS,t}}
&= 
P_{\text{TS}} H_{\text{TS,t}}(E^{(2)}_{-}-H_{\text{D}}-H_{\text{TS,L}})^{-1}(1-P_{\text{TS}}) H_{\text{TS,t}}\, P_{\text{TS}} ,
\end{split}
\end{equation}
where $P_{\text{TS}}=|1_{1},E^{(2)}_{-},0_{2}\rangle\langle1_{1},E^{(2)}_{-},0_{2}|+|0_{1},E^{(2)}_{-},1_{2}\rangle\langle0_{1},E^{(2)}_{-},1_{2}|$
is the projector on the $E^{(2)}_{-}$ state on the dot and the ground states of the TS leads. 
It acts within the reduced Hilbert space of the states $E^{(2)}_{\pm}, E^{(1)}_{\lambda}$ 
on the dot and the odd parity ground state subspace of the TS leads. 
In particular, $0_{\eta}$ ($1_{\eta}$) denotes the ground state in which the non-local fermionic mode
in TS $\eta$ is unoccupied (occupied). When evaluating Eq.~\eqref{Eq21} we find that 
the result is of the form as given in the main text by Eq.~(6) with $\nu=\text{TS}$ and
\begin{align}
\label{Eq24}
E^{0}_{\text{TS}}
&=
g_{\text{TS}}\left[
B_{\lambda\ua} 
T_{-}
\left(
A_{\lambda\ua} 
T_{-}
+
B_{\lambda\da} 
S_{-}
\right)
\left(
t_{1a\ua}
t_{2b\ua}
-
t_{1b\ua}
t_{2a\ua}
\right)
\right.
\left.
+
B^{2}_{\lambda\ua} 
S_{-}
T_{-}
\left(
t_{1b\da}t_{2a\ua}-t_{1a\ua}t_{2b\da}
\right)
\right]
\\
E^{a}_{\text{TS}}
&=
-
g_{\text{TS}}
\left[
\left(
A_{\lambda\ua}
T_{-}
+
B_{\lambda\da}
S_{-}
\right)^{2}
t_{1b\ua}t_{2b\ua}
+
B^{2}_{\lambda\ua}
\left(
S^{2}_{-}
t_{1b\da}t_{2b\da}
+
T^{2}_{-}
t_{1a\ua}t_{2a\ua}
\right)
-
B_{\lambda\ua}
S_{-}
\left(
A_{\lambda\ua}
T_{-}
+
B_{\lambda\da}
S_{-}
\right)
\left(
t_{1b\ua}
t_{2b\da}
+
t_{1b\da}
t_{2b\ua}
\right)
\right]
\nonumber,
\end{align}
where we have introduced the coefficient
\begin{equation}
g_{\text{TS}}
=
\frac{
2
}{
E^{(1)}_{\lambda}-E^{(2)}_{-}
}>0.
\end{equation}
There are also processes which do not transport a non-local fermion across the JJ 
and thus lead to a contribution $\widetilde{E}_{\text{TS}}$ which is independent of the 
superconducting phase difference. In these processes each TS interacts seperately with the QD. 
In particular this means that the action of the effective tunneling Hamiltonian on 
the two odd parity ground states of the TS is identical. Consequently, this contribution
is proportional to the identity operator and is not relevant when computing the zero-temperature 
Josephson current. 
For the case when $\theta=\pi/2$ we have listed all the intermediate tunneling sequences
which contribute to the Josephson current
in Fig.~\ref{supp:fig3}. 
The phase shift $\varphi^{0}_{\text{TS}}(\theta=\pi/2)$ for $\lambda=1,4$ is plotted
as a function of the external Zeeman field in Fig.~\ref{supp:fig7}.
Lastly, the phase shifts $\varphi^{0}_{\text{TS}}(\theta)$ and Josephson currents $I_{\text{TS}}(\theta)$
 at $\varphi_{\text{TS}}=0$ are plotted in Fig.~\ref{supp:fig6}.

\begin{figure}[t]  \centering
\includegraphics[width=0.35\linewidth] {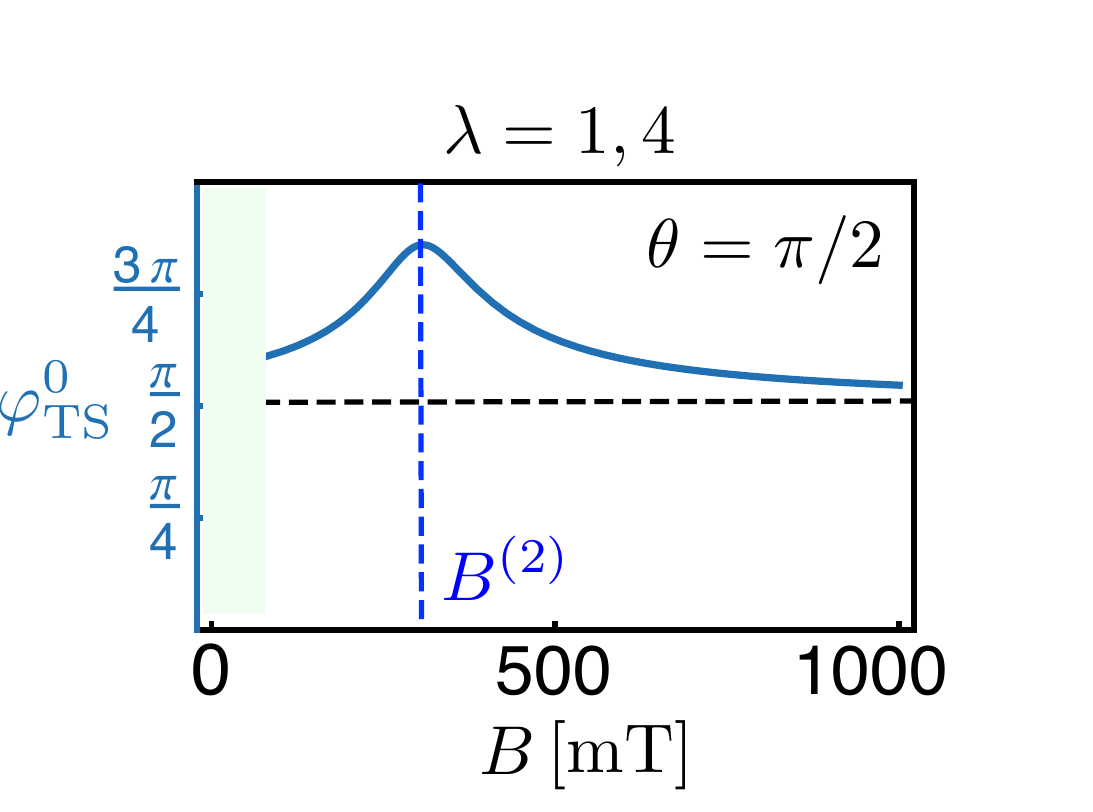}
\caption{
Phase shift $\varphi^{0}_{\text{TS}}$ as a function of the magnitude of the external 
magnetic field $B$ at $\theta=\pi/2$ for $\lambda=1,4$. For $\lambda=2,3$ the phase
shift is independent of $B$ and given by $\varphi^{0}_{\text{TS}}=\pi/2$.  
For the SC JJ we do not observe a phase shift when $\theta=\pi/2$, $\varphi^{0}_{\text{S}}=0$. 
We see that the phase shift is peaked at  $B=B^{(2)}$ when the singlet triplet mixing is maximal
and it saturates at $\pi/2$ when $B\gg B^{(2)}$. Note however that our
perturbative approach is not valid when $B\ll B^{(2)}$, because additional energy levels would have
to be taken into account. 
}\label{supp:fig7}
\end{figure}

\section{Critical angle}

\begin{figure}[t]  \centering
\includegraphics[width=0.5\linewidth] {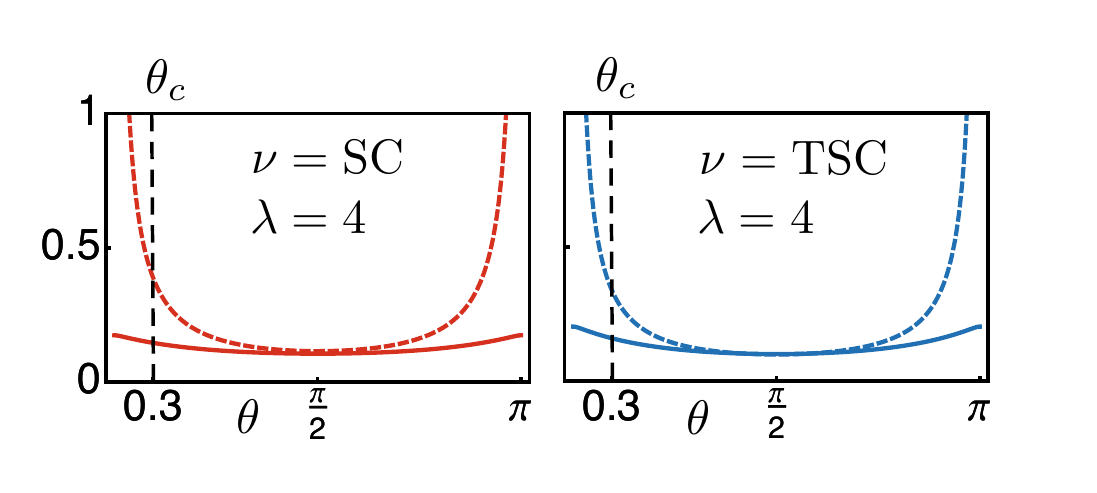}
\caption{
Estimate of the critical angle $\theta_{c}$ when $\lambda=4$ by analyzing the conditions 
for the weak coupling limit as a function of $\theta$. The system parameters are chosen 
as in the main text and supplemental material. 
In the left panel we plot $\pi\nu_{F}t^{2}/(\Omega\sin\theta)$ (red dashed) and 
$\pi\nu_{F}t^{2}/|E^{(1)}_{4}-E^{(2)}_{-}|$ (red solid). In the right panel we plot $t/(\Omega\sin\theta)$
(blue dashed) and $t/|E^{(1)}_{4}-E^{(2)}_{-}|$ (blue solid).
We find that $\theta_{c}=0.3$. This
choice of critical angle also works for $\lambda=1,2,3$.
}\label{supp:fig4}
\end{figure}

The effective Hamiltonians for the SC JJ and the TS JJ
are valid in the weak tunnel coupling limit. For the SC JJ this limit is defined by
\begin{equation}
\pi\nu_{F}t_{\eta\tau}t_{\eta'\tau'}\ll E^{(1)}_{\lambda}-E^{(2)}_{-},\Omega\sin(\theta),\Delta
\end{equation}
and for the TS JJ by
\begin{equation}
t_{\eta\tau}\ll E^{(1)}_{\lambda}-E^{(2)}_{-},\Omega\sin(\theta).
\end{equation}
These conditions fix a critical angle $\theta_{c}>0$ so that our perturbative approach 
is valid when $\theta\in[\theta_{c},\pi-\theta_{c}]$. In this section we want
to determine this critical angle for the system parameters which we have chosen in Fig.~3 of the main text. 
To get a sense of scales, we consider an InAs nanowire QD JJ with SC leads of length $L=1 \ \mu\text{m}$.
We assume that the effective mass of the electrons in the wire is given by $m= 0.05m_{e}$ where
$m_{e}$ is the bare electron mass. Furthermore, we expect that the Fermi energy of
the leads is given by $E_{F}=0.1$ meV and the induced superconducting gap by $\Delta=0.1$ meV. 
The density of states at the Fermi level of the nanowires in the normal metal state is given by 
$\nu_{F}=\frac{L}{\pi}\sqrt{\frac{2m}{\hbar^{2}}}\frac{1}{\sqrt{E_{F}}}$. 
For the order of magnitude of the tunnel coupling between dot and leads we assume that $t=0.01$ meV.
Furthermore, we fix $V_{g}$ so that $E^{(1)}_{\lambda}(\pi/2)-E^{(2)}_{-}(\pi/2)\approx0.1$ meV. 
This means that depending on the choice of $\lambda$ we have
$(\left.V_{g}\right|_{\lambda=1}, \left.V_{g}\right|_{\lambda=2}, \ \left.V_{g}\right|_{\lambda=4}, \ \left.V_{g}\right|_{\lambda=4})=(0.89 \ \text{meV},0.20 \ \text{meV},-0.12 \ \text{meV},-0.80 \ \text{meV})$. We can now graphically find an estimate for $\theta_{c}$, see Fig.~\ref{supp:fig4}. A choice of 
critical angle that works for all $\lambda$ is given by $\theta_{c}=0.3~.$

\section{Critical currents}

\subsection{Critical current of the SC JJ}

\begin{figure}[t]  \centering
\includegraphics[width=\linewidth] {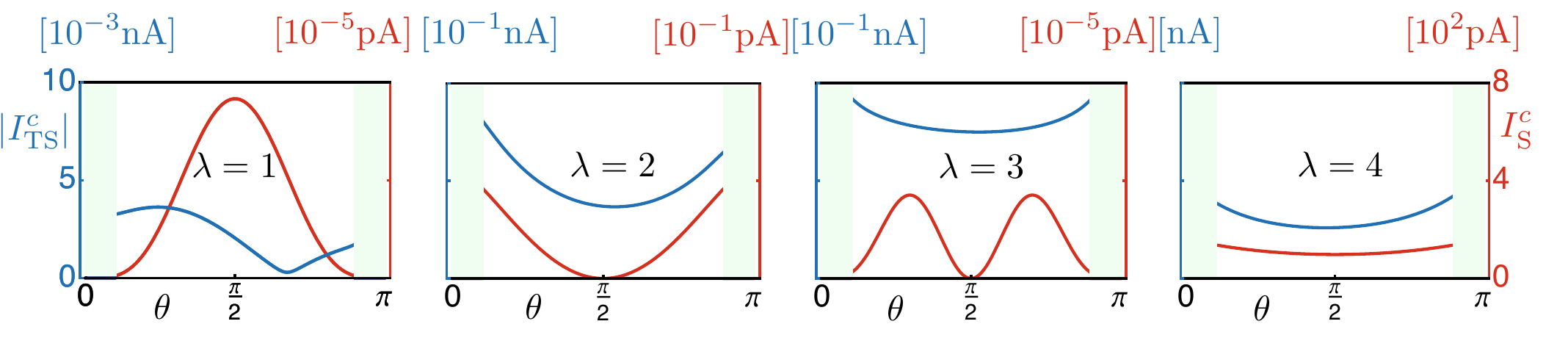}
\caption{
Magnitude of the critical current $|I^{c}_{\nu}(\theta)|$ for different choices of $\lambda$.
The system parameters are chosen as in the main text and supplemental material.
}\label{supp:fig5}
\end{figure}

In this section of supplemental material we compute the critical current $I_{\text{S},c}$.
First, we need to find an approximate value for the coefficient $g_{\text{S}}$. 
To this end, we notice that it can be rewritten as 
\begin{equation}
\begin{split}
g_{\text{S}}
&=
\frac{\Delta^{2}}{2}
\int^{\hbar\omega_{c}}_{-\hbar\omega_{c}}
\nu(E_{1}) \
\mathrm{d}E_{1}
\int^{\hbar\omega_{c}}_{-\hbar\omega_{c}}
\nu(E_{2}) \
\mathrm{d}E_{2}
\frac{
1
}{
\sqrt{E^{2}_{1}+\Delta^{2}}\sqrt{E^{2}_{2}+\Delta^{2}}
}
\\
&\times
\frac{
1
}{
\left[(E^{(1)}_{\lambda_{0}}-E^{(2)}_{-})+\sqrt{E^{2}_{1}+\Delta^{2}}\right]
\left[(E^{(1)}_{\lambda_{0}}-E^{(2)}_{-})+\sqrt{E^{2}_{2}+\Delta^{2}}\right]
\left[(E^{(2)}_{+}-E^{(2)}_{-})+\sqrt{E^{2}_{1}+\Delta^{2}}+\sqrt{E^{2}_{2}+\Delta^{2}}
\right]
}
\end{split}
\end{equation}
where $\nu(E)=\sum_{\bk}\delta(E-E_{\bk})$ is the density of state of the leads in the normal state  at energy $E$ and $\omega_{c}$ is a cut-off frequency
which is typically of the order of the Debye frequency of the crystal.
For simplicity, we now assume that $\nu(E)\approx\nu_{F}$ for $|E|\geq\Delta$ and $\nu(E)=0$ for $|E|<\Delta$. 
This yields
\begin{equation}
\begin{split}
g_{\text{S}}
&\approx
\frac{(\Delta\nu_{F})^{2}}{2}
\left(
\int_{-\hbar\omega_{c}}^{-\Delta}
\mathrm{d}E_{1}
+
\int^{\hbar\omega_{c}}_{\Delta}
\mathrm{d}E_{1}
\right)
\left(
\int_{-\hbar\omega_{c}}^{-\Delta}
\mathrm{d}E_{2}
+
\int^{\hbar\omega_{c}}_{\Delta}
\mathrm{d}E_{2}
\right)
\frac{
1
}{
\sqrt{E^{2}_{1}+\Delta^{2}}\sqrt{E^{2}_{2}+\Delta^{2}}
}
\\
&\times
\frac{
1
}{
\left[(E^{(1)}_{\lambda_{0}}-E^{(2)}_{-})+\sqrt{E^{2}_{1}+\Delta^{2}}\right]
\left[(E^{(1)}_{\lambda_{0}}-E^{(2)}_{-})+\sqrt{E^{2}_{2}+\Delta^{2}}\right]
\left[(E^{(2)}_{+}-E^{(2)}_{-})+\sqrt{E^{2}_{1}+\Delta^{2}}+\sqrt{E^{2}_{2}+\Delta^{2}}
\right]
}.
\end{split}
\end{equation}
Defining $\xi_{\pm}=(E^{(1)}_{\lambda}-E^{(2)}_{\pm})/\Delta$ allows us to rewrite 
this expression as 
\begin{equation}
\begin{split}
g_{\text{S}}
&\approx
\frac{4\alpha}{\pi^{2}}
\frac{m L^{2}}{\hbar^{2}\Delta E_{F}}.
\end{split}
\end{equation}
where we have introduced the dimensionless factor 
\begin{equation}
\alpha=\int^{\infty}_{1}
\mathrm{d}x
\int^{\infty}_{1}
\mathrm{d}y \ 
\frac{
1
}{
\sqrt{1+x^{2}}\sqrt{1+y^{2}}
\left(\sqrt{1+x^{2}}+\sqrt{1+y^{2}}+\xi_{-}-\xi_{+}
\right)
\left(\sqrt{1+x^{2}}+\xi_{-}\right)
\left(\sqrt{1+y^{2}}+\xi_{-}\right)
}
\end{equation}
and we have assumed that $\hbar\omega_{c}\gg\Delta$ which ensures that the Cooper potential of the BCS theory is a good approximation to the actual electron pairing potential.
We note that $\alpha$ is a function of the relative orientation of SOI axis and Zeeman field,
$\alpha=\alpha(\theta)$. For the system parameters chosen in the main text we find that
$\alpha\approx10^{-1}$.
In total the critical current is then given by
\begin{equation}
I_{\text{S}}^{c}\approx
\frac{8\alpha}{\pi^{2}}
\frac{meL^{2}}{\hbar^{3}\Delta E_{F}}
\sqrt{\left(E^{0}_{\text{S}}\right)^{2}+\left(E^{a}_{\text{S}}\right)^{2}}\text{sgn}(E^{0}_{\text{S}}).
\end{equation}
We have plotted $I^{c}_{\text{S}}(\theta)$ in Fig.~\ref{supp:fig5}. 
For the case when $\theta=\pi/2$ and $\lambda=2,3$ we have $I_{\text{S}}^{c}=0$ because $B_{2(3)\ua}=0$. 
Moreover, there exists a significant difference in magnitude of the critical currents
for the cases when $\lambda=1,4$ which are most relevant for our experimental proposal in the main text. 
We can understand this because 
$\left.I_{\text{S},c}\right|_{\lambda=1}/\left.I_{\text{S},c}\right|_{\lambda=4}\propto(B_{1 \ua}/B_{4 \ua})^{4}\approx10^{-6}$: 
The virtual state $E^{(1)}_{1}$ only contains a small amount of $B_{1 \ua}$ due to the SOI, while 
$E^{(1)}_{4}$ consists mostly of $B_{4 \ua}$, hence $B_{4 \ua}\gg B_{1 \ua}$.
The conclusion is that the absence or presence of a phase shift can most easily be measured
when virtual tunneling occurs via the $E^{(1)}_{4}$ state. 

\subsection{Critical current of the TS JJ}
For the TS JJ we find that the critical current is given by 
\begin{equation}
I_{\text{TS}}^{c}=\frac{4\kappa_{\text{TS}}e
}{\hbar
(E^{(1)}_{\lambda_{0}}-E^{(2)}_{-})
}\sqrt{\left(E^{0}_{\text{TS}}\right)^{2}+\left(E^{a}_{\text{TS}}\right)^{2}}\text{sgn}(E^{0}_{\text{TS}}).
\end{equation}
We plot $I^{c}_{\text{TS}}(\theta)$ in Fig.~\ref{supp:fig5}. 
Again we see a significant difference in magnitude when comparing the most relevant cases of $\lambda=1$
and $\lambda=4$. This can be explained in the same way as for the SC JJ. 
However, this time we have for example at $\theta=\pi/2$, $\left.I_{\text{TS},c}\right|_{\lambda=1}/\left.I_{\text{TS},c}\right|_{\lambda=4}\propto(B_{1 \ua}/B_{4 \ua})^{2}\approx10^{-3}$. 

\begin{figure}[t]  \centering
\includegraphics[width=0.75\linewidth] {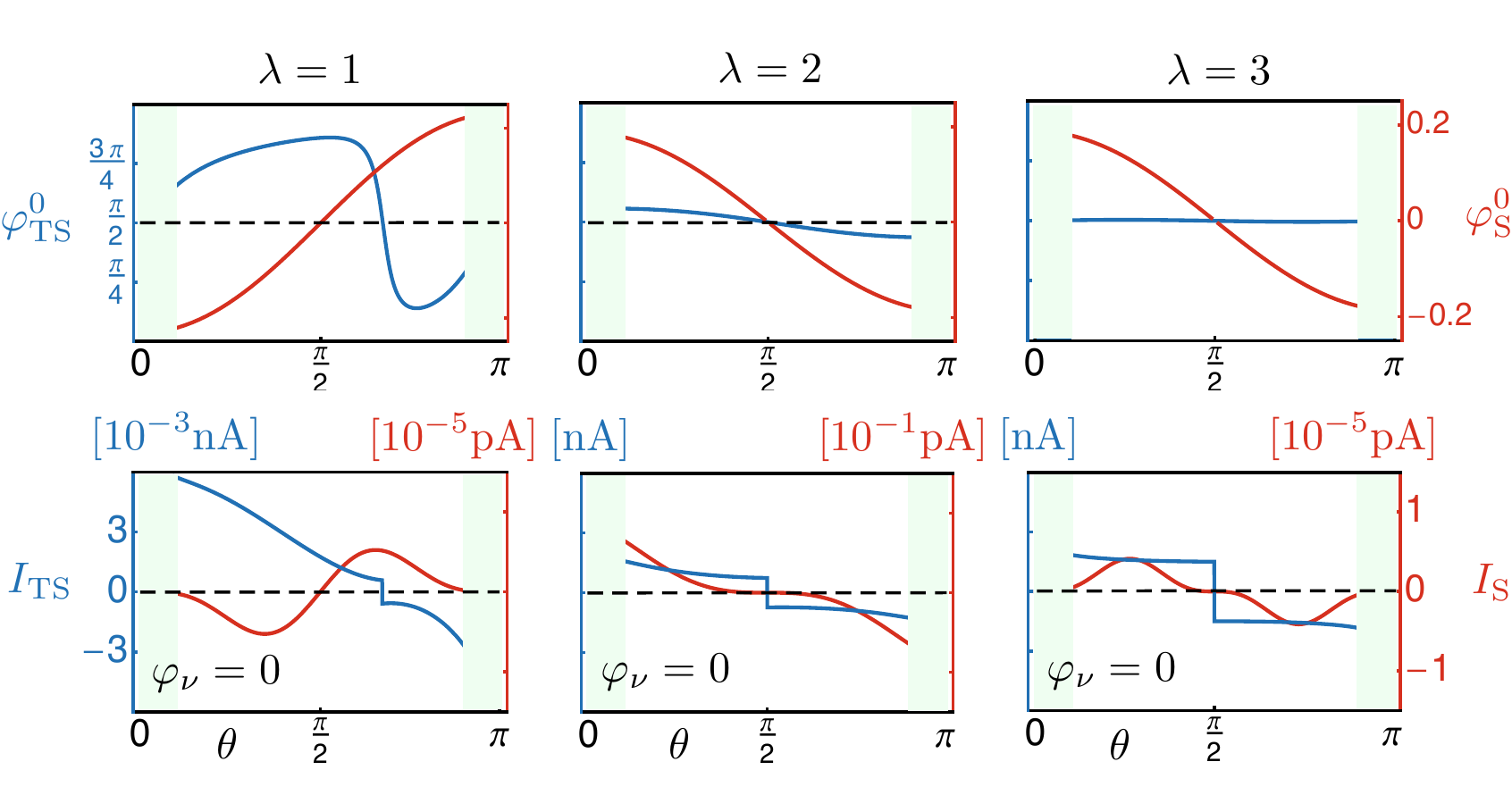}
\caption{
Phase shift $\varphi^{0}_{\nu}(\theta)$ (top row) and Josephson current $I_{\nu}(\theta)$ at $\varphi_{\text{S}}=0$ (bottom row) for $\lambda=1,2,3$. The system parameters are chosen as
in the main text. The jumps in the Josephson current $I_{\text{TS}}(\theta)$ correspond to
a change of the ground state of the junction.
}\label{supp:fig6}
\end{figure}

\end{widetext}

\end{document}